\documentclass[showpacs,preprintnumbers,amsmath,amssymb,floatfix,nofootinbib]{revtex4}

\usepackage{graphicx}
\usepackage{dcolumn}
\usepackage{bm}

\usepackage{pstricks}

\newcommand{\sherpa}{{\tt SHERPA}}

\begin{document}
\title{Simulating $W/Z$+jets production at the CERN LHC}
 \homepage{http://www.physik.tu-dresden.de/~krauss/hep/}

\author{Frank Krauss}
\email{krauss@theory.phy.tu-dresden.de}
\affiliation{Institute for Theoretical Physics, D-01062 Dresden, Germany}
\author{Andreas Sch{\"a}licke}
 \email{dreas@theory.phy.tu-dresden.de}
\affiliation{Institute for Theoretical Physics, D-01062 Dresden, Germany}
\author{Steffen Schumann}
 \email{steffen@theory.phy.tu-dresden.de}
\affiliation{Institute for Theoretical Physics, D-01062 Dresden, Germany}
\author{Gerhard Soff\footnote{deceased}} 
\affiliation{Institute for Theoretical Physics, D-01062 Dresden, Germany}

\begin{abstract}
\noindent
The merging procedure of tree-level matrix elements and the subsequent 
parton shower as implemented in the new event generator SHERPA will be 
validated for the example of single gauge boson production at the LHC.
The validation includes consistency checks and comparisons to results 
obtained from other event generators. In particular, comparisons with 
full next-to-leading order QCD calculations prove SHERPA's ability to 
correctly account for additional hard QCD radiation present in these 
processes.
\end{abstract}

\maketitle

\psset{linewidth=2pt}
\psset{unit=1pt}

\section{Introduction}

\noindent
The production of electroweak gauge bosons, which decay leptonically, 
is one of the most prominent examples for hard processes at hadron colliders 
and one of the first applications of perturbative QCD in such reactions. In 
fact, the next-to-leading order (NLO) corrections to this process in QCD, 
calculated by \cite{Altarelli:1979ub,Kubar-Andre:1978uy,Harada:1979bj,
Abad:1978ke,Humpert:1979ux} provided the first calculation of such corrections 
for hadron collisions. Later, their production cross section has been 
calculated at NNLO by \cite{Hamberg:1990np,Harlander:2002wh}. Recently, 
the first distribution determined at NNLO related to these processes, 
namely the boson rapidity, has been calculated by \cite{Anastasiou:2003ds}. 
In addition, there is a large number of computer programs dealing with single 
gauge boson production. They range from {\tt RESBOS} \cite{Balazs:1997xd}, which 
resums soft gluon effects in these processes, to codes, that evaluate cross 
sections at the LO level for the production of gauge bosons accompanied by jets. 
Examples for the latter include specialised ones, such as {\tt VECBOS} 
\cite{Berends:1990ax}, and general ones, usually called parton level generators, 
such as {\tt COMPHEP} \cite{Pukhov:1999gg}, {\tt GRACE/GR@PPA} 
\cite{Ishikawa:1993qr,GRAPPA}, {\tt MADGRAPH/MADEVENT} 
\cite{Stelzer:1994ta,Maltoni:2002qb}, {\tt ALPGEN} \cite{Mangano:2002ea}, 
and {\tt AMEGIC++} \cite{Krauss:2001iv}. Furthermore, the first package called 
{\tt MCFM} has been made available that calculates total and differential cross 
sections at NLO precision for the production of gauge bosons with up to two jets 
\cite{Campbell:2002tg,Campbell:2003hd}. This reflects the importance of this
particular process. At the LHC becoming operational in the near future, the gauge 
bosons will be produced with unprecedented rates. For instance, at luminosities of 
${\cal L}\approx 10^{33}{\rm cm}^2/{\rm s}$ the production and leptonic decay 
of a single $W$ boson will occur with a frequency of around 20 Hz, rendering 
this process a prime candidate for luminosity monitoring 
\cite{Khoze:2000db,Dittmar:1997md,Martin:1999ww,Giele:2001ms}. Of course, these 
large rates will allow to measure the gauge bosons parameters, such as their 
masses and widths, with a precision \cite{ATLAS_TDR,Haywood:1999qg} beyond what 
could be reached at previous collider experiments \cite{Abachi:1995xc,
Affolder:2000bp,Abazov:2002bu,Abazov:2003sv,Albajar:1990hg,Alitti:1991dm,
Abe:1995wf,Abbott:1999tt,Affolder:2000mt,Abazov:2002xj}. At the CERN LHC, the
production of $W$ and $Z$ bosons together with jets also constitute an important
background to all kinds of searches for new physics, such as supersymmetry. 
An example for this is the production and decay of gluinos, where the production of
jets plus a $Z$ boson decaying into neutrinos forms a major background.

\noindent
In a previous analysis \cite{Krauss:2004bs} it has been shown that some results 
for this type of process, i.e.\ the production of single gauge bosons plus extra jets, 
as obtained by other multi-purpose event generators such as {\tt PYTHIA} 
\cite{Sjostrand:1993yb,Sjostrand:2001yu}, {\tt HERWIG} \cite{Corcella:2000bw,Corcella:2002jc}
or even {\tt MC@NLO} \cite{Frixione:2002ik,Frixione:2003ei,Frixione:2004wy} differ 
significantly from the results obtained by \sherpa\ \cite{Gleisberg:2003xi}. In 
particular, it has been shown that already at the Fermilab Tevatron, operating
at roughly 2 TeV centre-of-mass energy, the additional jets are produced at 
significantly larger transverse momenta. The reason for this difference
is the way the different codes implement the knowledge of exact matrix elements 
for the production of multi-particle final states. In both, {\tt PYTHIA} and {\tt HERWIG},
they are included at first order in $\alpha_S$ through a correction of the first 
hard emission on the corresponding $q\bar q\to Vg$ or $qg\to Vq$ matrix element, 
where $V$ stands for the vector boson \cite{Seymour:1995df,Miu:1999ju,Corcella:2000gs}. 
In {\tt MC@NLO} the full first order correction, including both, the virtual and the real parts, 
are matched with the parton shower. This has the additional benefit that {\tt MC@NLO} 
reproduces correctly the total production rate of single gauge bosons and the spectrum 
of the first additional jet at first order in $\alpha_S$. In contrast, in \sherpa\ a
method has been implemented that consistently adds different matrix elements at the 
tree level for different jet multiplicities and merges them with 
the parton shower. The basic idea in this approach is to internally define a region 
of jet production (hard parton emissions) and a region of jet evolution (soft parton emissions).
The two regimes are divided by a $k_\perp$-type of jet measure \cite{Catani:1991hj,
Catani:1992zp,Catani:1993hr}. Leading higher order effects are incorporated by reweighting 
the matrix elements with appropriate Sudakov form factors. Formal independence at 
leading logarithmic order of the overall result on the jet measure is achieved 
by suitable starting conditions and vetoing hard emissions inside the parton shower. 
This approach was presented for the first time \cite{Catani:2001cc} for $e^+e^-$ collisions; 
it has been extended to hadronic collisions in \cite{Krauss:2002up}. A reformulation for 
dipole cascades has been presented in \cite{Lonnblad:2001iq}. The algorithm is implemented 
in a fully automated way and in full generality in \sherpa, some other realisations 
\cite{Mrenna:2003if,Mangano:2001xp} proved the flexibility and the validity of the approach.

\noindent
In this publication, the previous analysis \cite{Krauss:2004bs} will be extended to
the case of the CERN LHC, operating in the $pp$ mode at 14 TeV centre-of-mass energy. 
In Sec.\ \ref{Consistency_Sec}, a number of consistency checks will focus on the
independence of the results on variations of the internal jet definition and of the number 
of matrix elements involved. Also, the effect of scale variations in both the matrix elements
and the parton shower is investigated there. Then, in Sec.\ \ref{Compare_NLO_Sec}, 
results obtained with \sherpa\ will be contrasted to those obtained from fixed order (LO and NLO) 
calculations provided by {\tt MCFM}. Following this, different multi-purpose event generators,
namely {\tt PYTHIA}, {\tt MC@NLO} and \sherpa, will be compared in Sec.\ \ref{Compare_MC_Sec}, 
before the conclusions will summarise the findings. 

\section{Consistency checks}\label{Consistency_Sec}

\noindent
Before comparing the results of \sherpa\ with those of other programs, some consistency
checks will be performed. To do so, the dependence of some observables in reactions of
the type $pp \to e^+e^- +X$ on internal parameters intrinsic for the merging 
procedure will be investigated. In particular, these parameters are the internal 
jet resolution cut $Q_{\rm cut}$ and the maximal number $n_{\rm max}$ of final state 
partons (giving rise to jets) described through matrix elements. The former parameter 
defines the transition of the matrix element domain to the phase space region covered by 
the parton shower during event generation. In principle, the actual value of this parameter 
can be chosen freely, nevertheless their exist criteria that guide such a choice. For 
very low values of $Q_{\rm cut}$ the evaluation of the matrix elements becomes very 
challenging and potentially inefficient once jet cuts are performed on the analysis 
level harder than the generation cut $Q_{\rm cut}$. The upper limit is defined by the 
scale where jets produced by the parton shower start to disagree significantly
from such produced by equivalent matrix elements. To study especially the effect of the 
upper limit, the values used in this analysis will range over nearly one order of magnitude, 
from 15 to 100 GeV. The choice of the number of matrix element partons taken into account 
may be steered by two aspects. First of all, $n_{\rm max}$ has to be sufficiently large to 
properly account for the phase space region the observable under consideration is 
sensitive to. As an example, consider the transverse momentum of the boson
compared to that of the, say, third jet. It is obvious that a rather inclusive 
quantity such as the former may be appropriately described with lower values 
of $n_{\rm max}$ than the latter observable. On the other hand, the upper limit on
$n_{\rm max}$ is given by the availability of the matrix elements at all and by 
the potentially large amount of CPU time the evaluation of multi-leg matrix elements 
requires. Within \sherpa\ matrix elements with up to four extra partons can be delivered
for the processes under consideration in this publication. After evaluating the sensitivity of 
the results on the principal parameters defining the merging procedure, $Q_{\rm cut}$ and
$n_{\rm max}$, the effect of scale variations will be investigated. This, together with
the dependence on $Q_{\rm cut}$ and $n_{\rm max}$ yields an estimate for the uncertainty 
related to predictions of \sherpa.

\noindent
The results presented in this section were generated with the following setups: when varying 
the jet resolution parameter $Q_{\rm cut}$, the maximal number of final state partons 
$n_{\rm max}$ has been set to $n_{\rm max}=3$. When studying the impact of different matrix 
element multiplicities, the scale $Q_{\rm cut}$ has been fixed to $Q_{\rm cut}=15$ GeV; 
this clearly maximises the impact of the higher order matrix elements. When scale variations
are under consideration, the choice $Q_{\rm cut}=20$ GeV and $n_{\rm max}=2$ has been made.

\noindent
In the following, $Z$-boson production will be investigated in more detail. Nevertheless, 
the process under consideration is $pp\to Z/\gamma^* \to e^+e^- +X$, where the full $\gamma$-$Z$ 
interference is taken into account and spin correlations are fully respected. Further 
input parameters used and the phase space cuts applied are summarised in the appendix 
\ref{appendix}. Note that the cut on the invariant mass of the lepton pair is just 
$m_{ee}> 15\,{\rm GeV}$ which is rather small. The description of such low mass lepton 
pairs constitutes a real challenge for the description through the merging prescription. The 
reason is that at large $Q_{\rm cut}={\cal O}(100\,\mbox{\rm GeV})$, lepton pairs with such 
low invariant mass clearly are softer than any jet produced through the matrix element, 
rendering a consistent merging a complicated task.
 
\subsection{Observables related to the leptons}

\begin{figure*}[t!]
\begin{center}
\begin{pspicture}(400,295)
\put(250,145){\includegraphics[width=5.5cm]{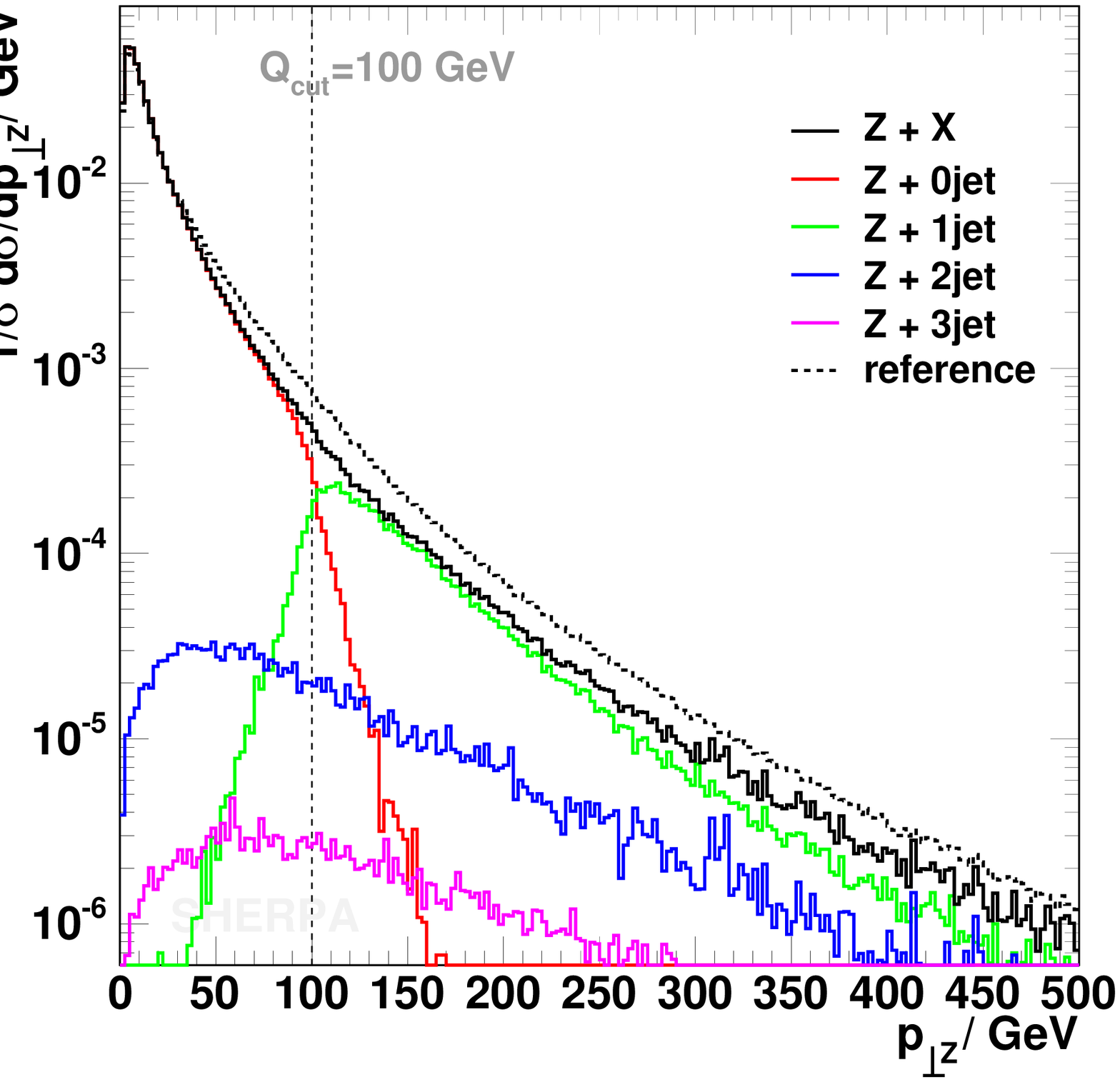}}
\put(125,145){\includegraphics[width=5.5cm]{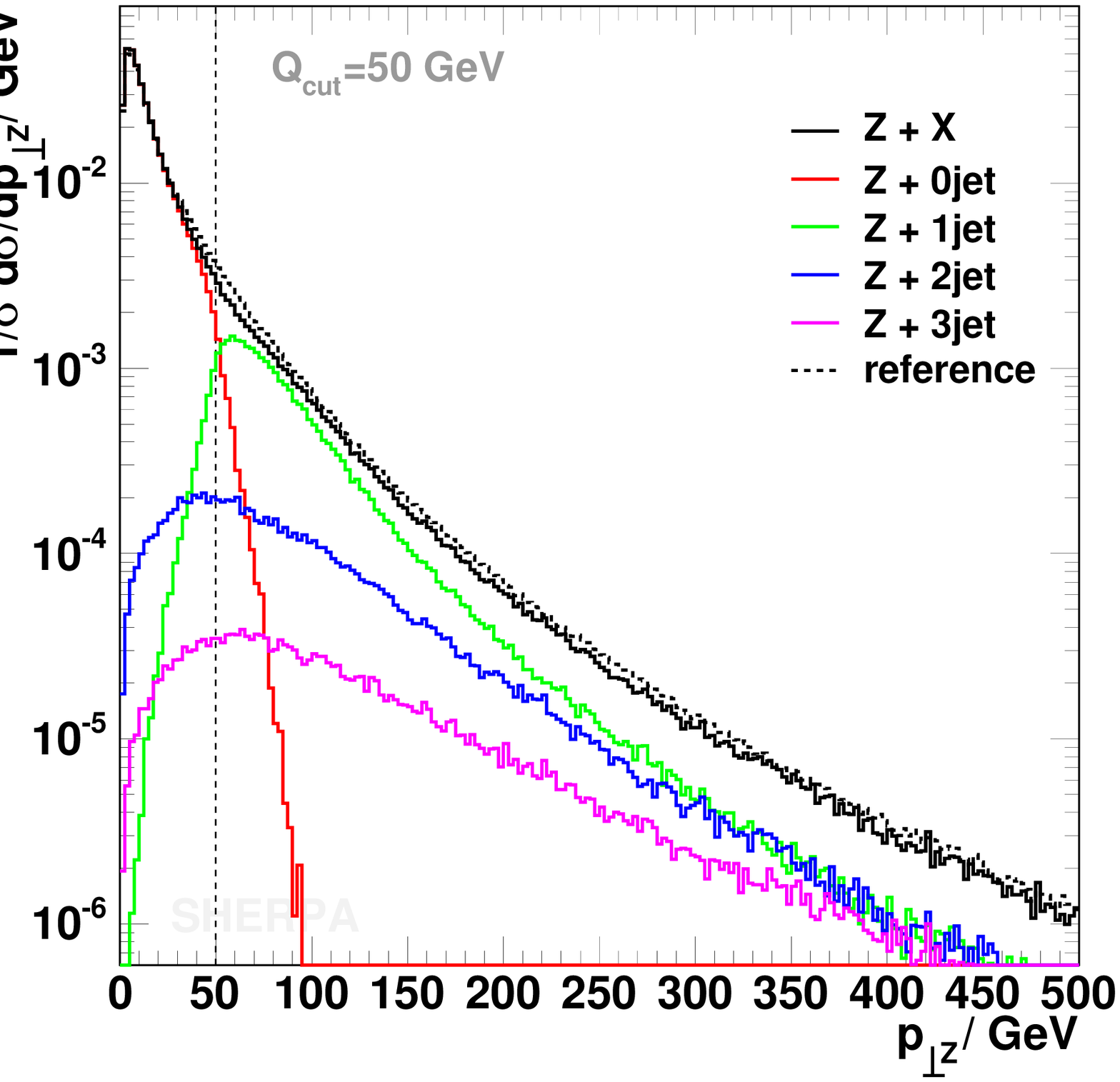}}
\put(0,145){\includegraphics[width=5.5cm]{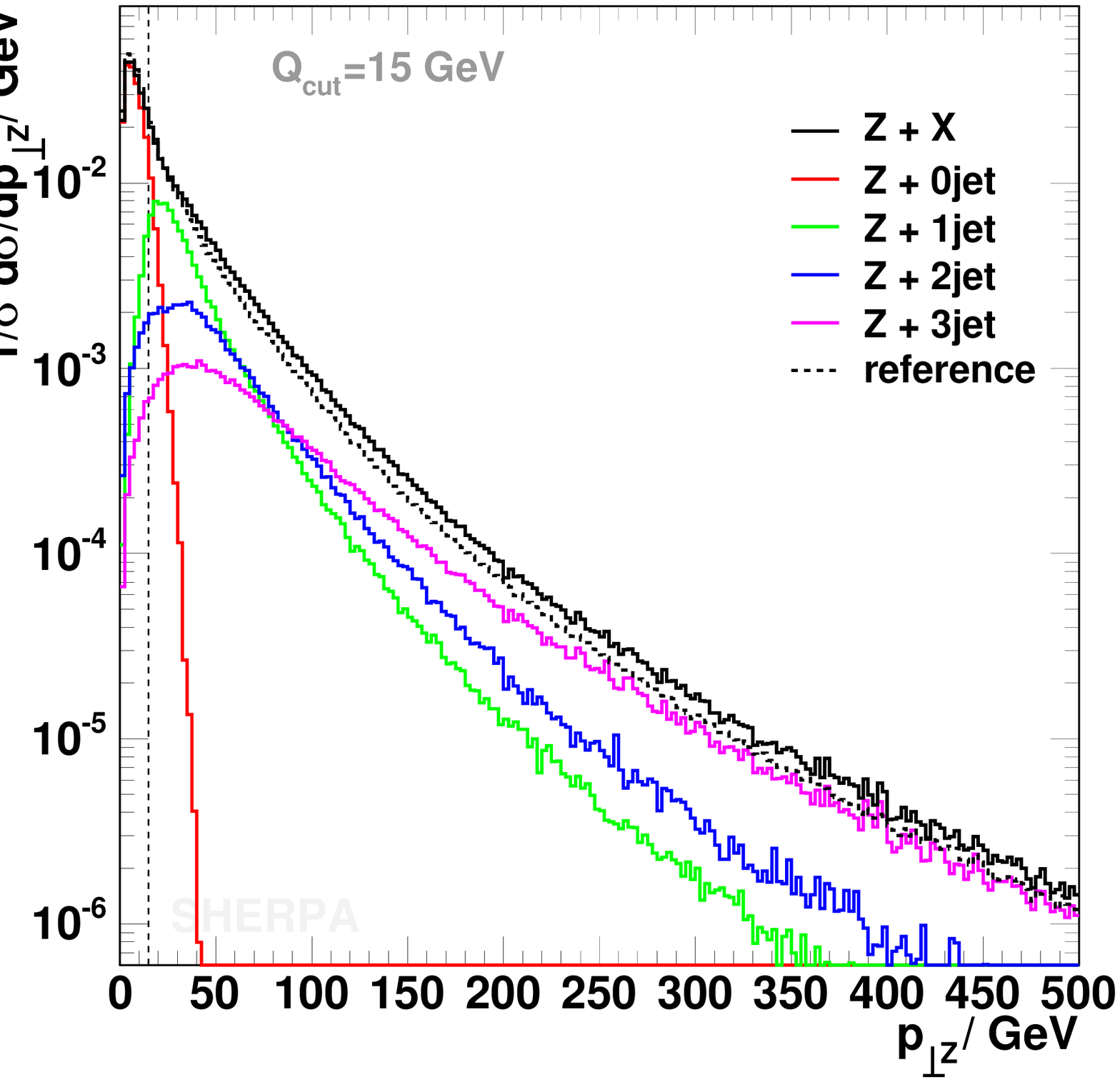}}
\put(250,0){\includegraphics[width=5.5cm]{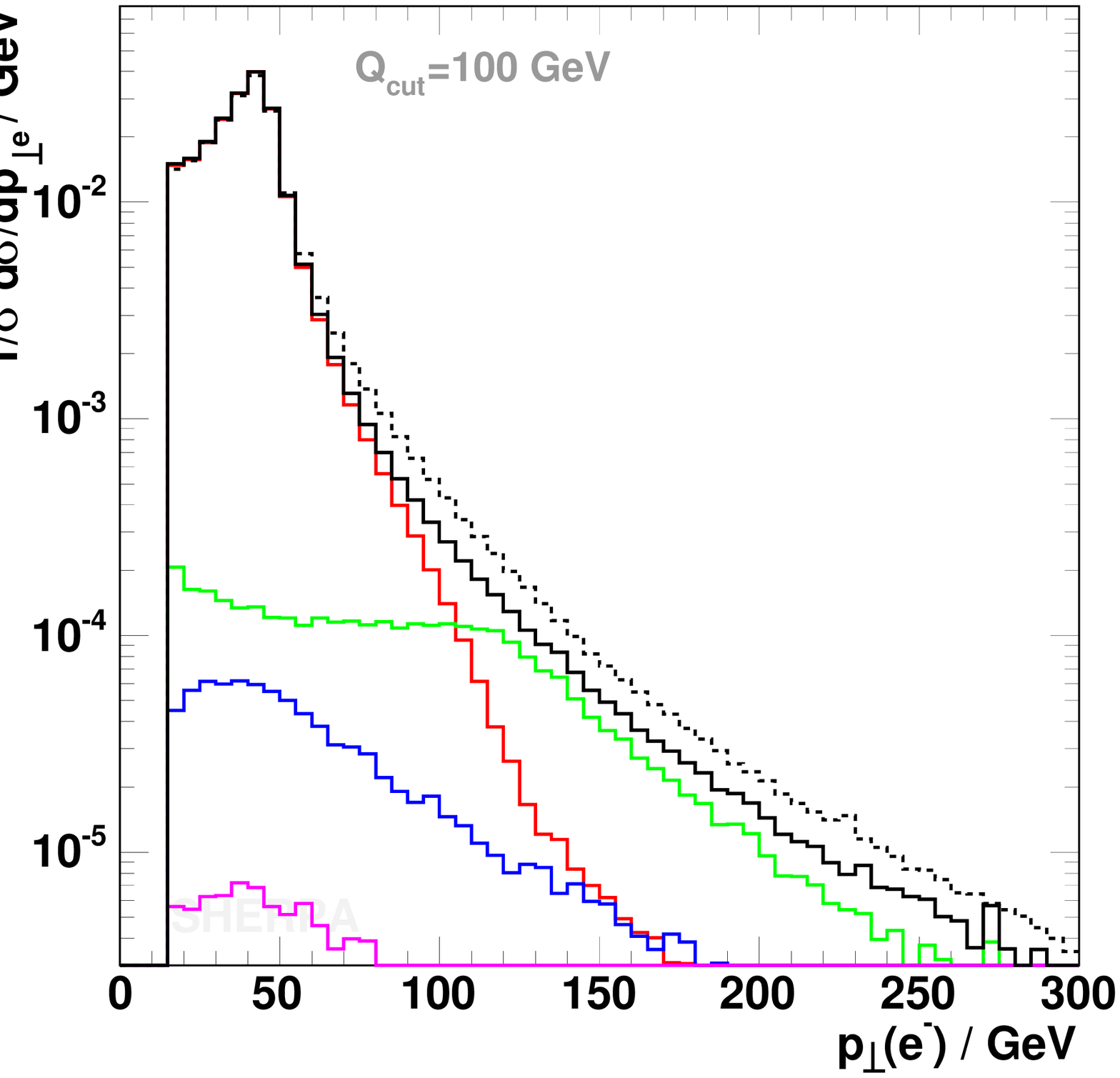}}
\put(125,0){\includegraphics[width=5.5cm]{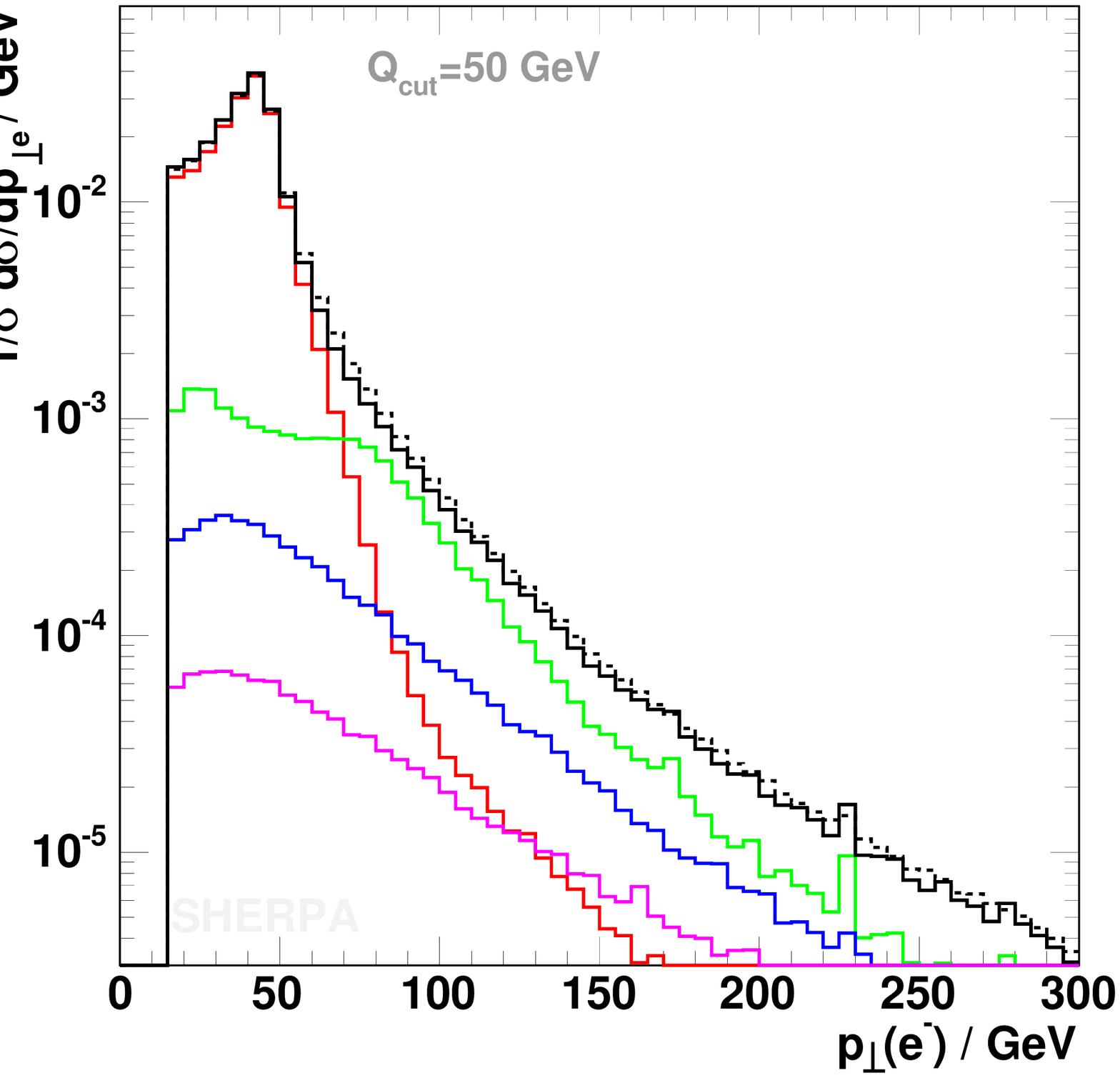}}
\put(0,0){\includegraphics[width=5.5cm]{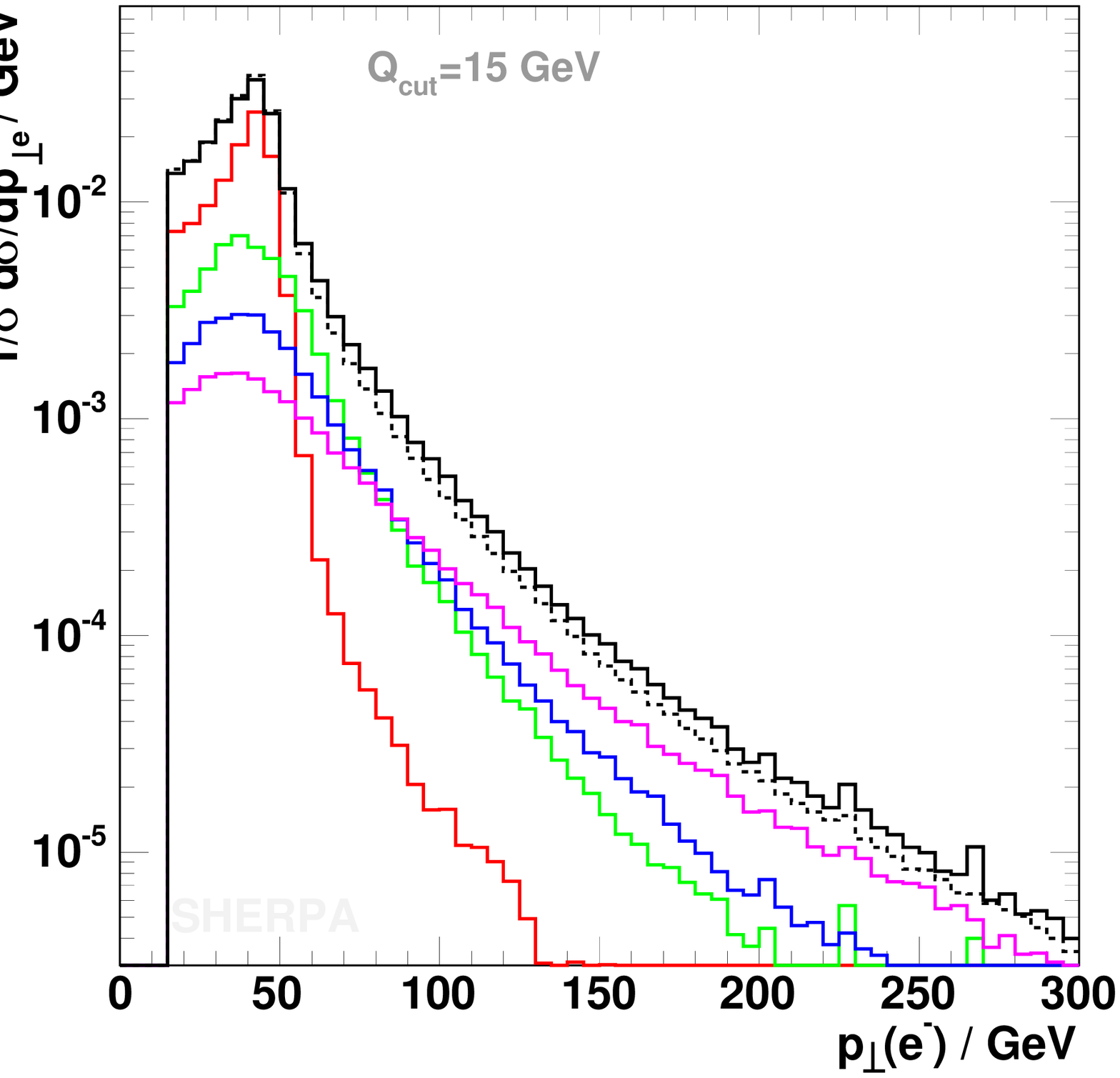}}
\end{pspicture}
\end{center}
\caption{\label{ycut_pt}$p_\perp(Z)$ (upper row) and $p_\perp(e^-)$ (lower row) for 
         $Q_{\rm cut}=15$ GeV, $50$ GeV and $100$ GeV (from left to right). 
         The dashed reference spectrum has been obtained after averaging the results 
         for $Q_{\rm cut}=15,\,20,\,30,\,50,\,100$ GeV.}
\end{figure*}

\noindent
Starting from more inclusive observables, first of all the effect of parameter variation on
lepton observables will be considered. In Fig.\ \ref{ycut_pt}, the $p_\perp$ spectra of both the
lepton pair (upper row) and of the electron alone (lower row) are shown for three different
values of $Q_{\rm cut}$: from left to right, in the columns $Q_{\rm cut} = 15,\,50,\,100$ GeV,
as indicated by the thin vertical lines. In each plot, the resulting spectrum is compared
to a reference obtained from averaging the results for $Q_{\rm cut} = 15,\,20,\,30,\,50,\,100$ GeV. 
In this and all other plots, contributions stemming from the different matrix element 
multiplicities are indicated through coloured lines. 

\noindent
Using $Q_{\rm cut}=15$ GeV obviously produces the hardest boson/lepton spectrum. 
It is the smallest cut considered here and therefore the distributions are dominated by 
matrix elements that in contrast to the parton shower favour rather hard parton kinematics. 
For very high $p_\perp$ the distributions are almost completely covered by the matrix element 
with the highest multiplicity ($n_{\rm max}=3$). This shows that the LHC provides 
enough phase space to produce a sufficient amount of events with three and more jets of 
$p_\perp>Q_{\rm cut}$. For the case of $Q_{\rm cut}=50$ GeV the situation is 
slightly different. The high-$p_\perp$ tail is filled to an equal amount by the different 
multiplicities, the total sum being slightly below the reference curve. This reference curve 
contains three results with jet resolutions smaller than $50$ GeV that somehow dominate the 
averaged result. The spectrum for $Q_{\rm cut}=100$ GeV starts to underestimate w.r.t.\ the
reference the boson transverse momentum at $p_\perp\approx 35$ GeV and the lepton $p_\perp$ 
for values larger than $60$ GeV. To understand this, one has to remember that the boson 
transverse momentum for values below the resolution cut is almost completely covered 
by the parton shower. The shower description, however, is known to suffer from a lack of 
hard QCD radiation. This leaves not enough hard partons, the boson can recoil against. 
Beyond this influence of the $Q_{\rm cut}$ variation on the intermediate and high boson 
transverse momenta, it has to be noted that all curves are very smooth around the jet 
resolution cut. Although the cut defines a rather sharp transition from the parton shower 
to the matrix element domain no significant holes in the boson and lepton $p_\perp$ spectra 
can be observed. 

\begin{figure*}[ht!]
\begin{center}
\begin{pspicture}(400,295)
\put(250,145){\includegraphics[width=5.5cm]{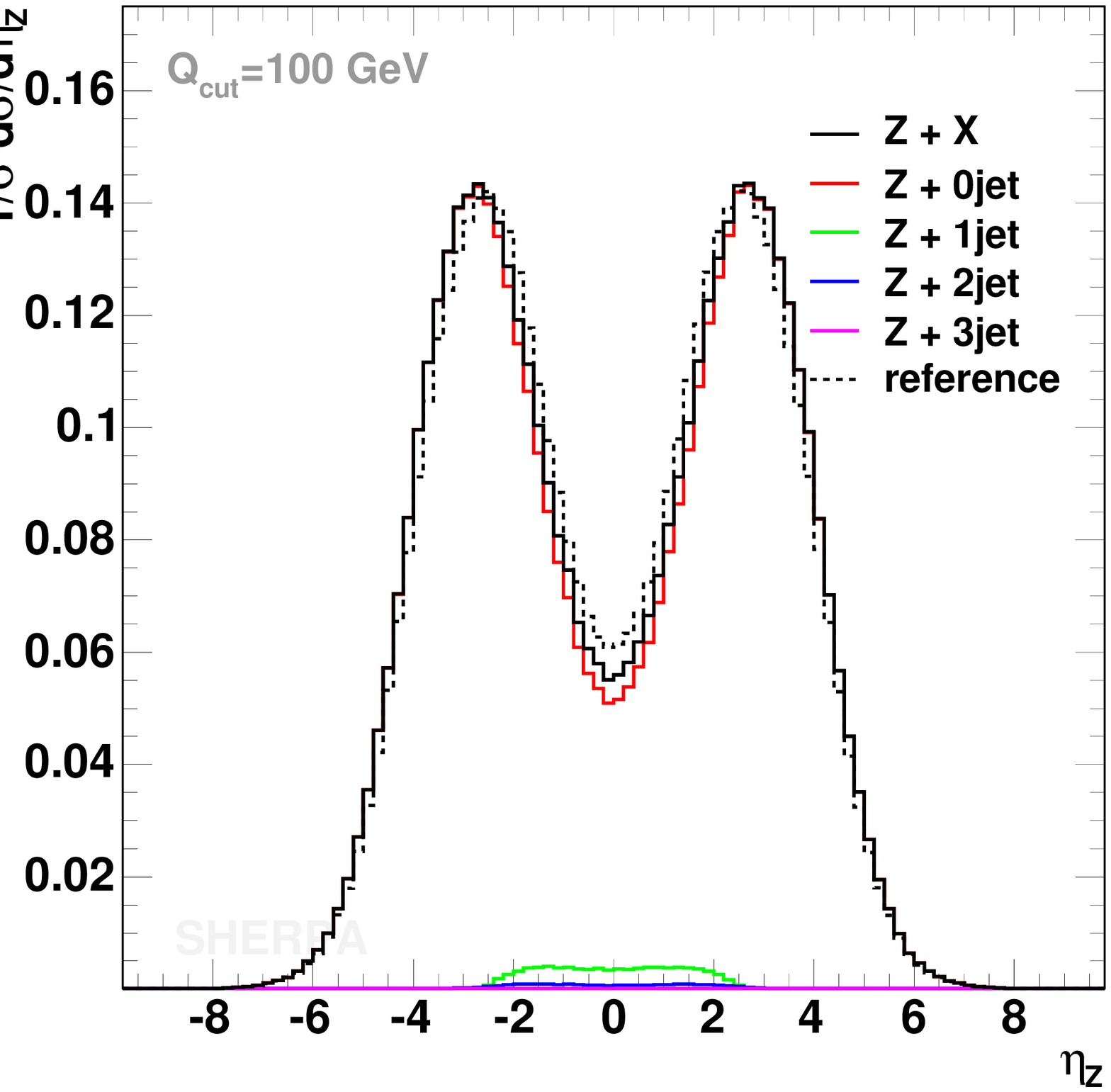}}
\put(125,145){\includegraphics[width=5.5cm]{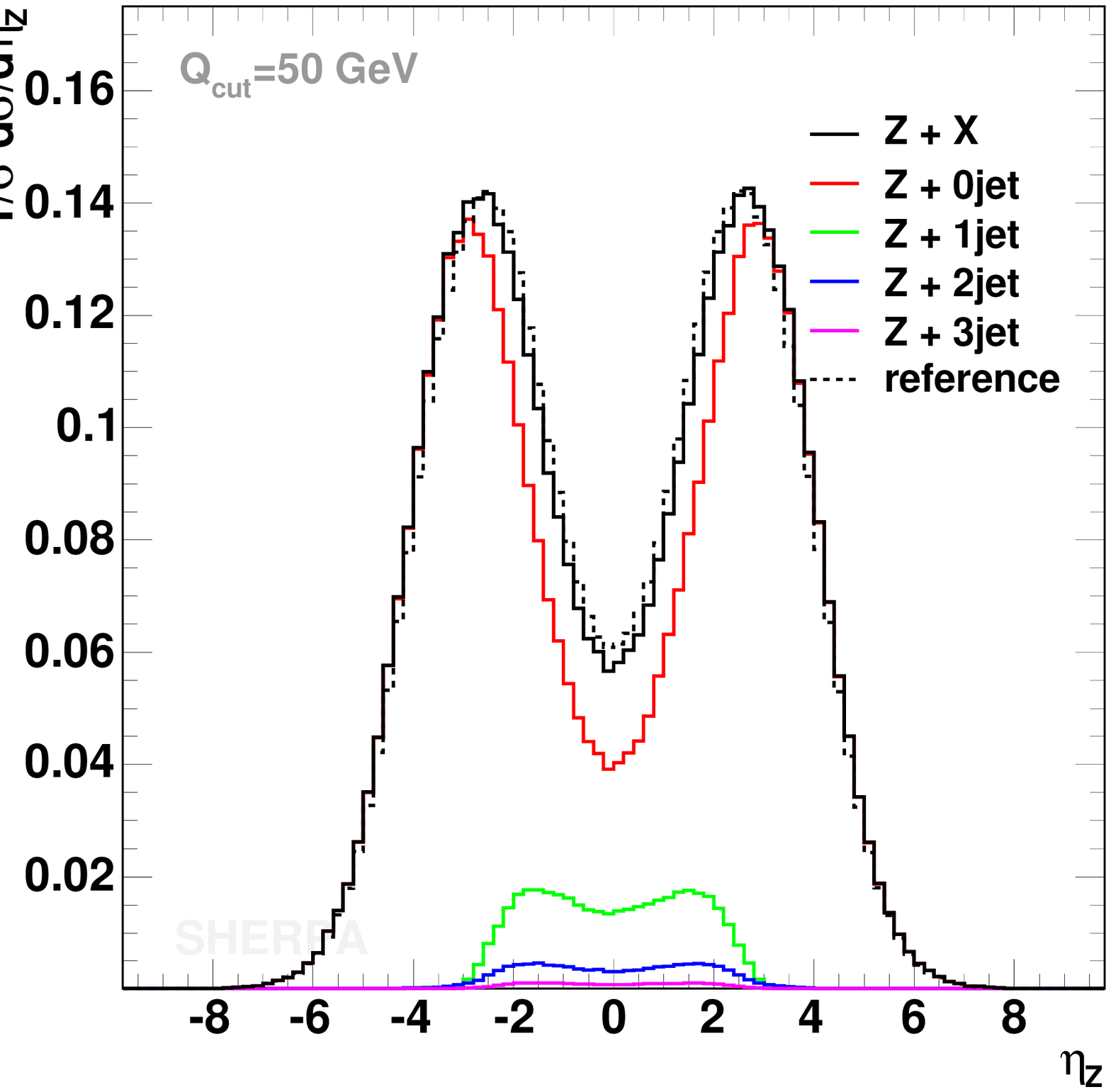}}
\put(0,145){\includegraphics[width=5.5cm]{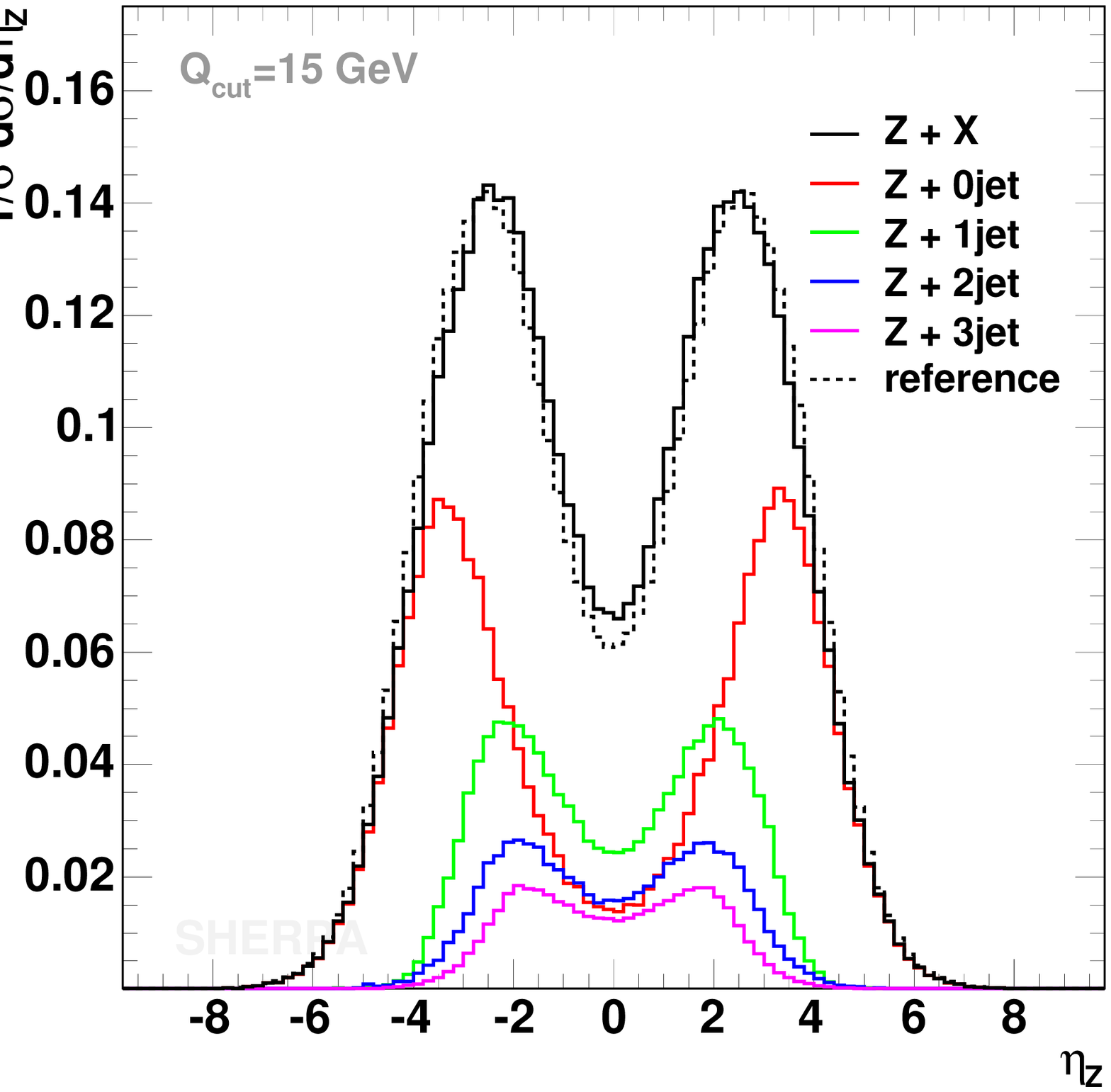}}
\put(250,0){\includegraphics[width=5.5cm]{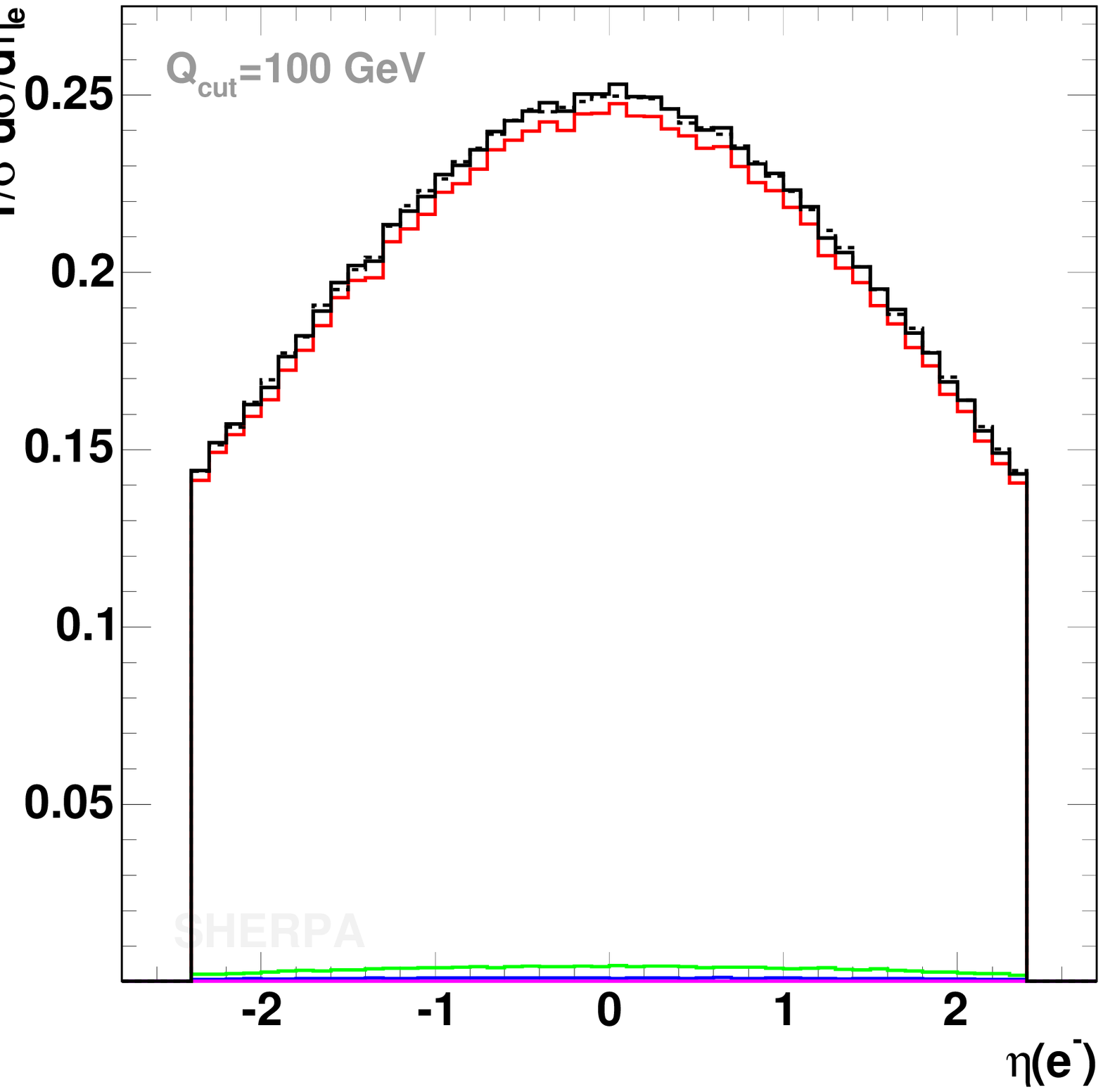}}
\put(125,0){\includegraphics[width=5.5cm]{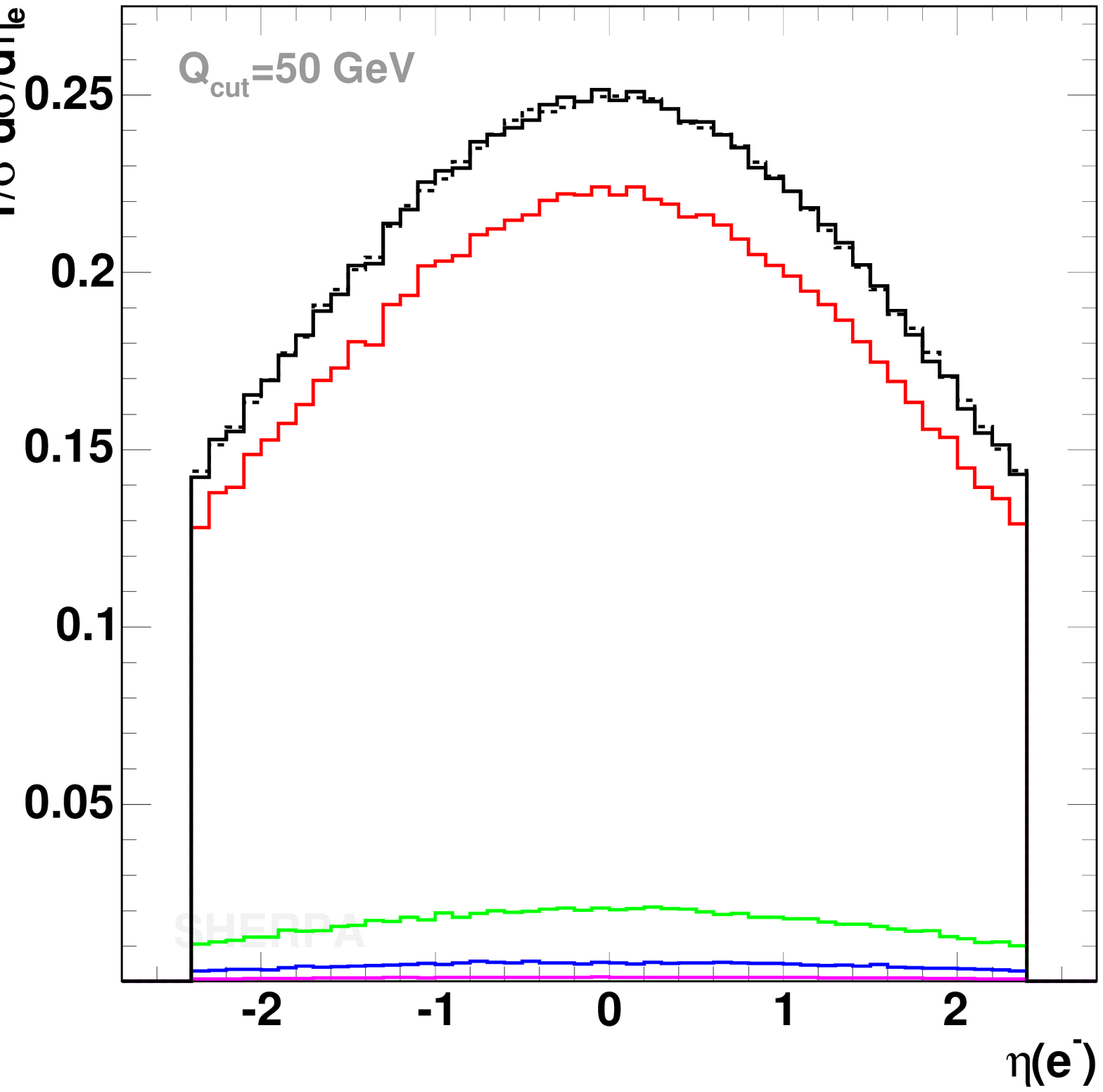}}
\put(0,0){\includegraphics[width=5.5cm]{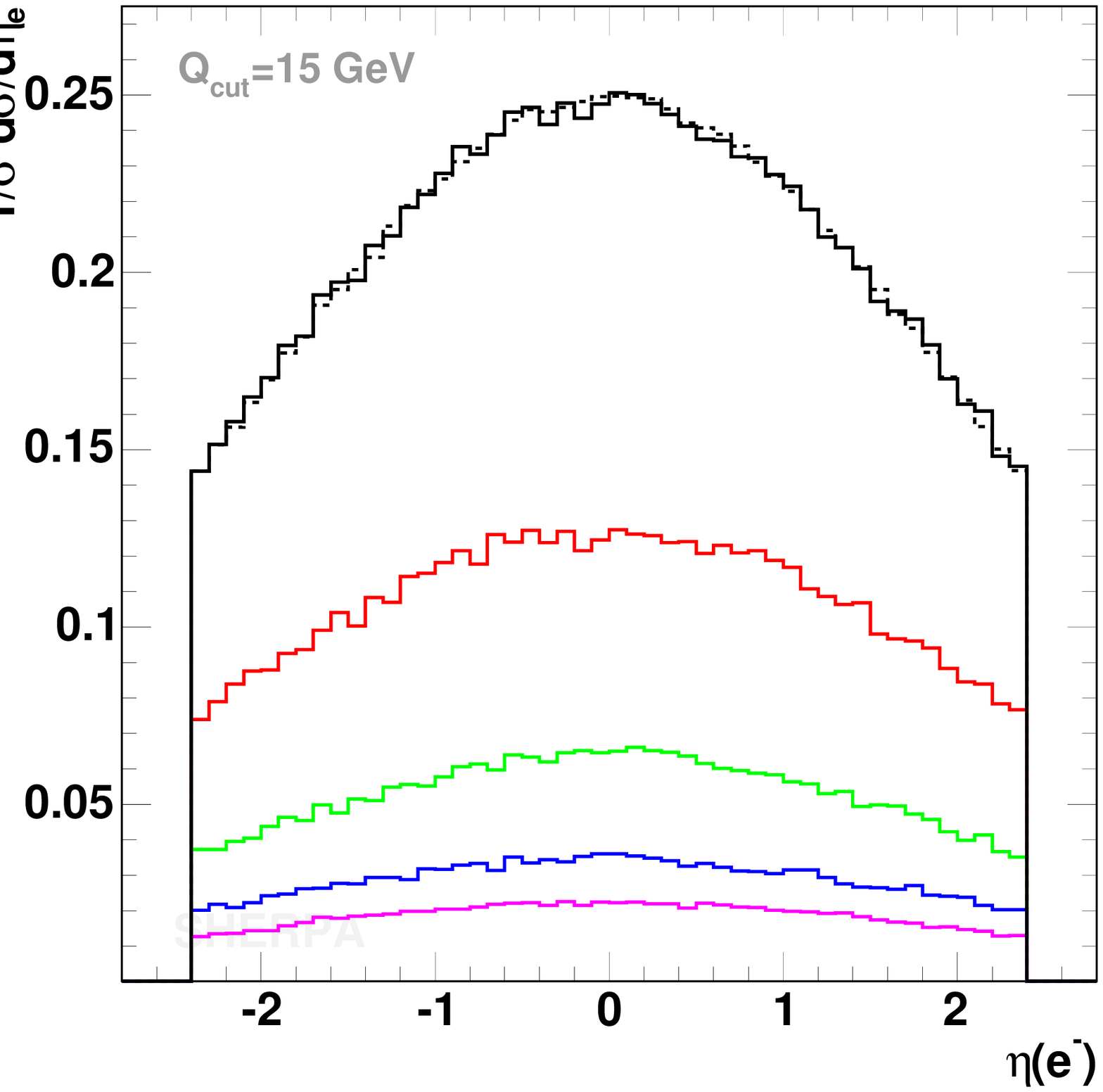}}
\end{pspicture}
\end{center}
\caption{\label{ycut_eta}$\eta(Z)$ (upper row) and $\eta(e^-)$ (lower row) for 
         $Q_{\rm cut}=15$ GeV, $50$ GeV and $100$ GeV (from left to right). 
         The dashed reference spectrum has been obtained after averaging the results 
         for $Q_{\rm cut}=15,\,20,\,30,\,50,\,100$ GeV.}
\end{figure*}

\noindent
In Fig.\ \ref{ycut_eta} the pseudo-rapidity spectra of the lepton pair and the single electron 
are displayed; again for $Q_{\rm cut}=15,\,50,\,100$ GeV with the same way of generating 
the reference. While the electron observable is nearly unaltered, the differences in the 
$\eta$ distribution of the lepton pair can be understood easily: the smaller the chosen cut, 
the larger the influence of the matrix elements with extra external legs. These matrix 
elements however prefer to produce the boson much more central than the parton shower does. 
This effect yields slightly tighter spectra with the central rapidities being pronounced 
for smaller resolution cuts. 

\begin{figure*}[h]
\begin{center}
\begin{pspicture}(400,295)
\put(250,145){\includegraphics[width=5.5cm]{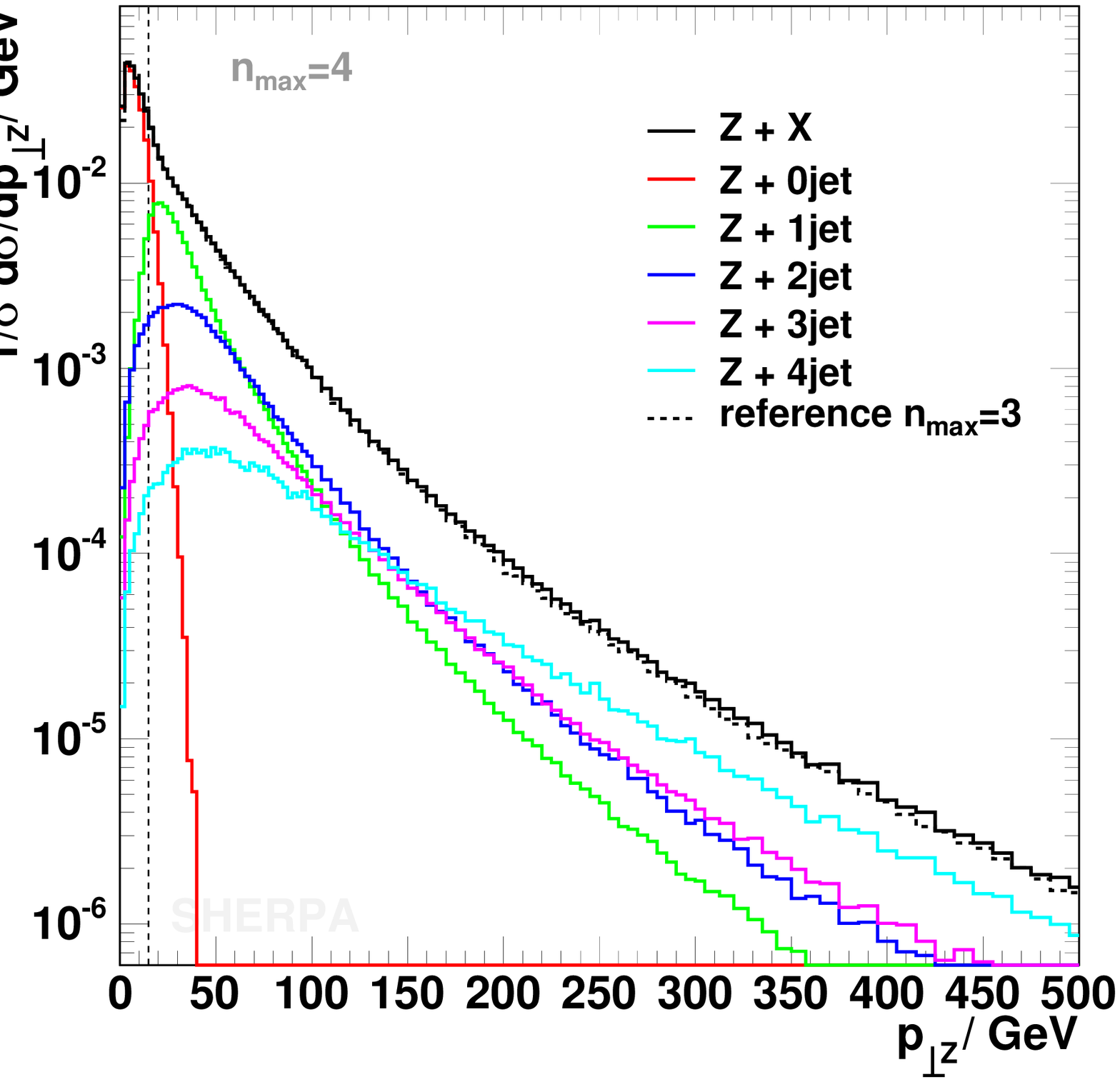}}
\put(125,145){\includegraphics[width=5.5cm]{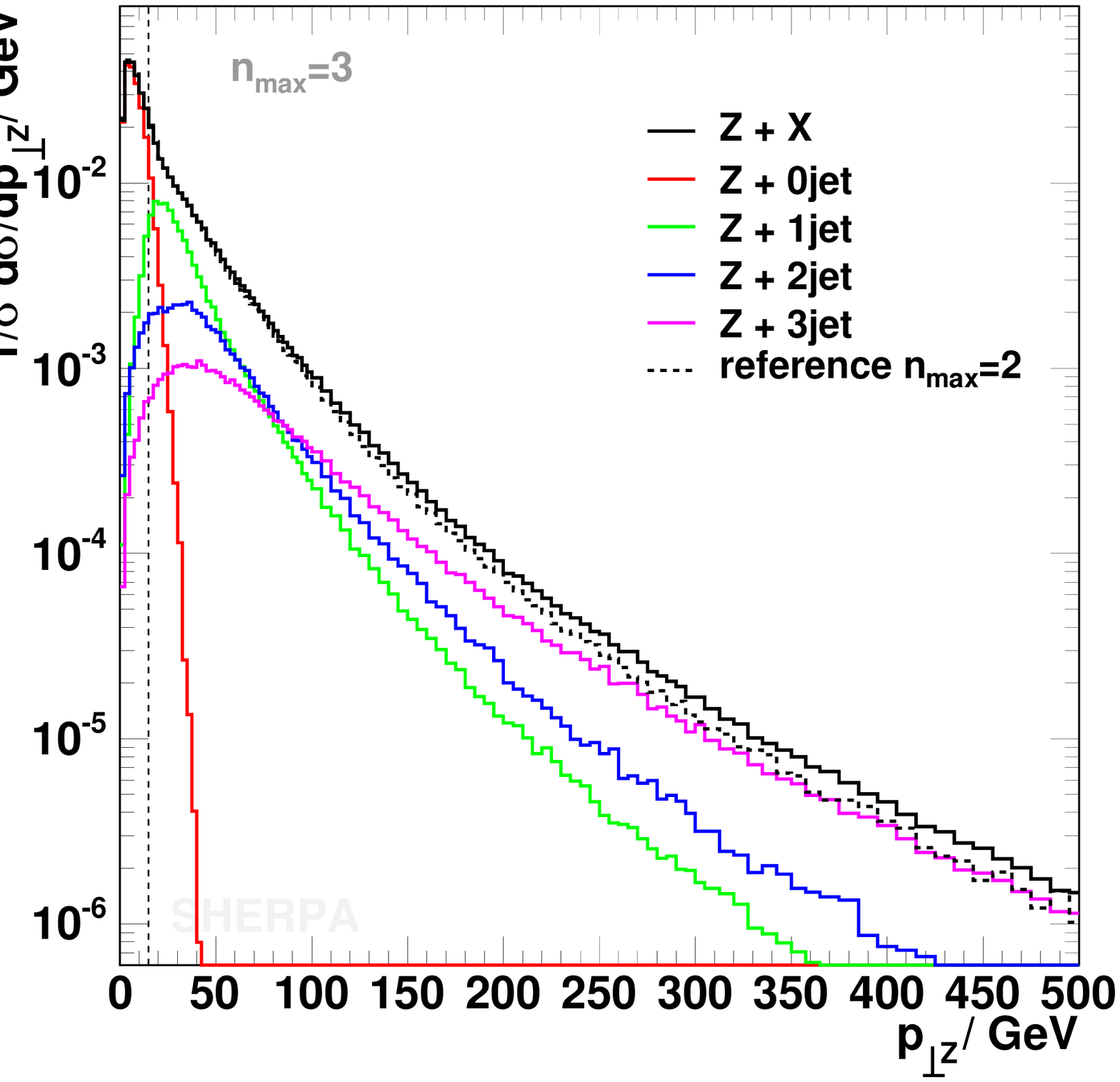}}
\put(0,145){\includegraphics[width=5.5cm]{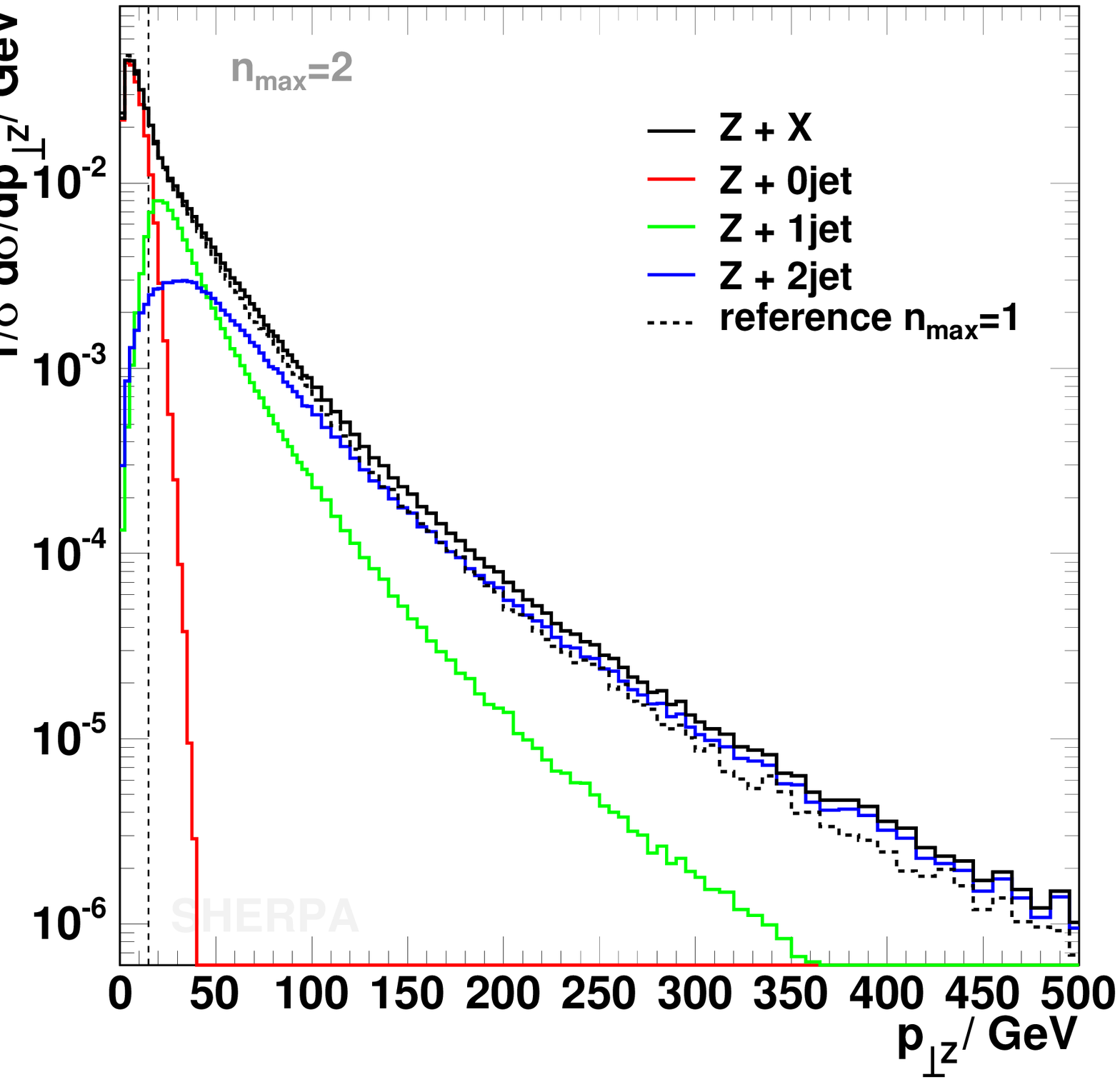}}
\put(250,0){\includegraphics[width=5.5cm]{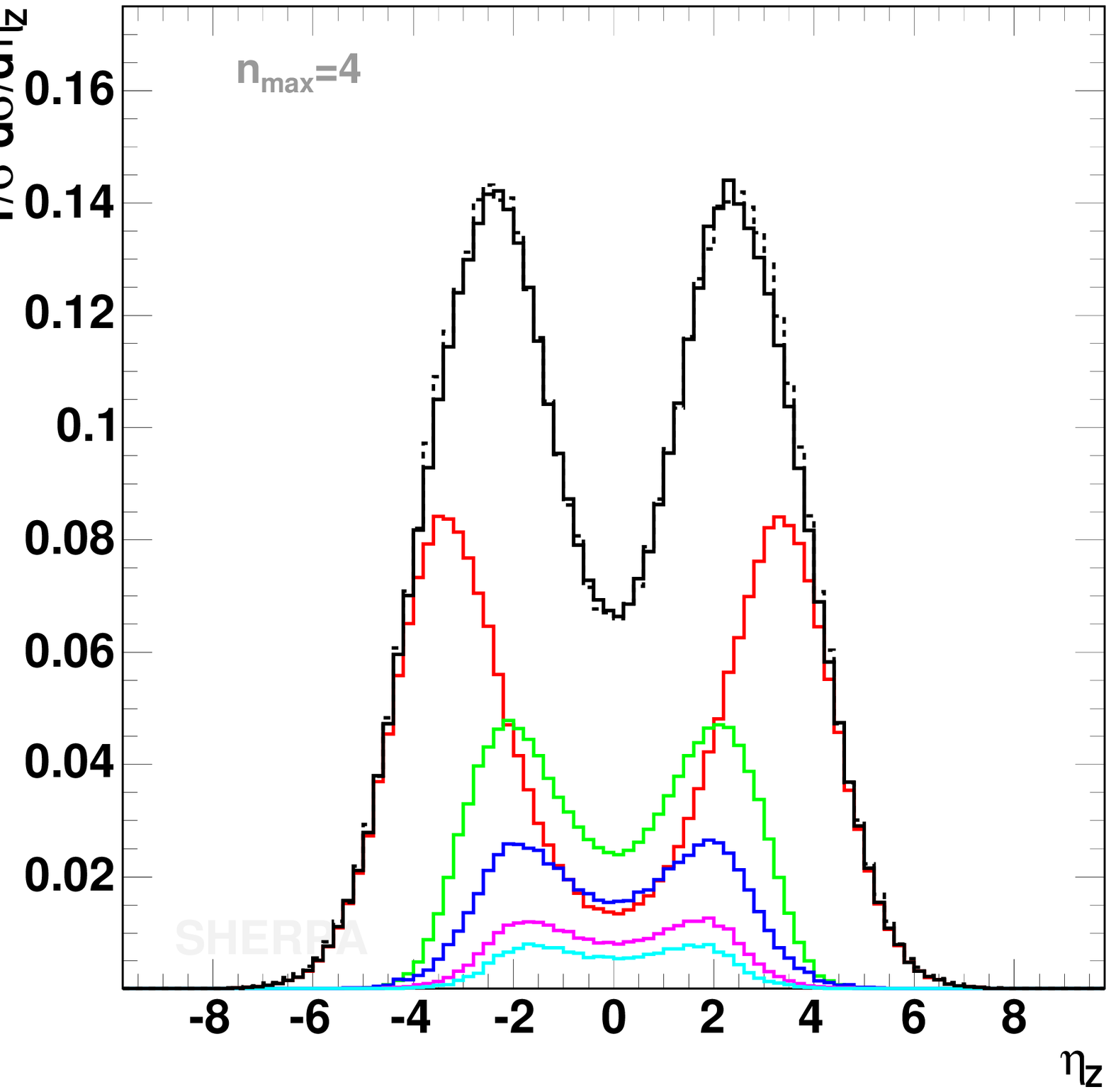}}
\put(125,0){\includegraphics[width=5.5cm]{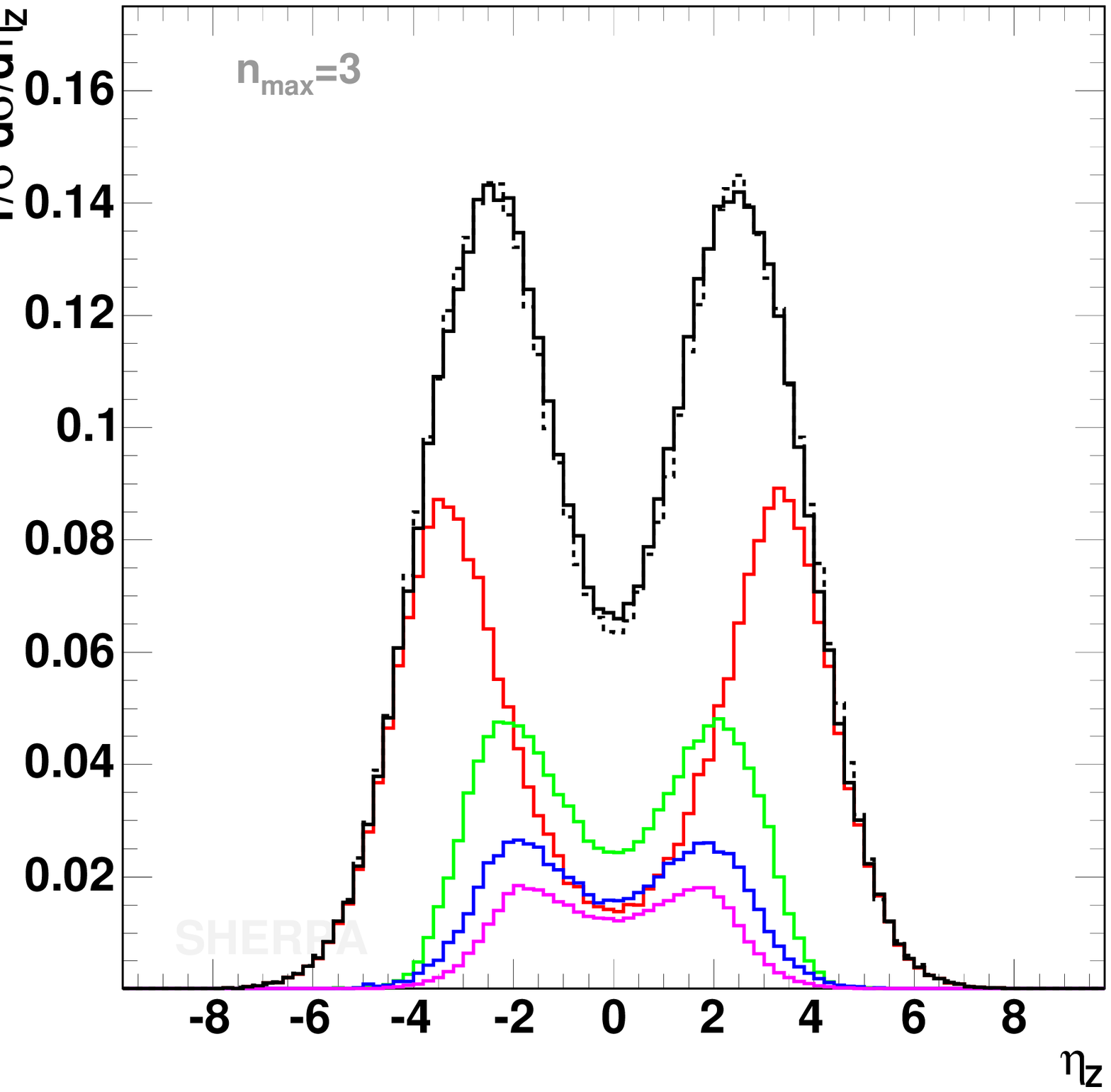}}
\put(0,0){\includegraphics[width=5.5cm]{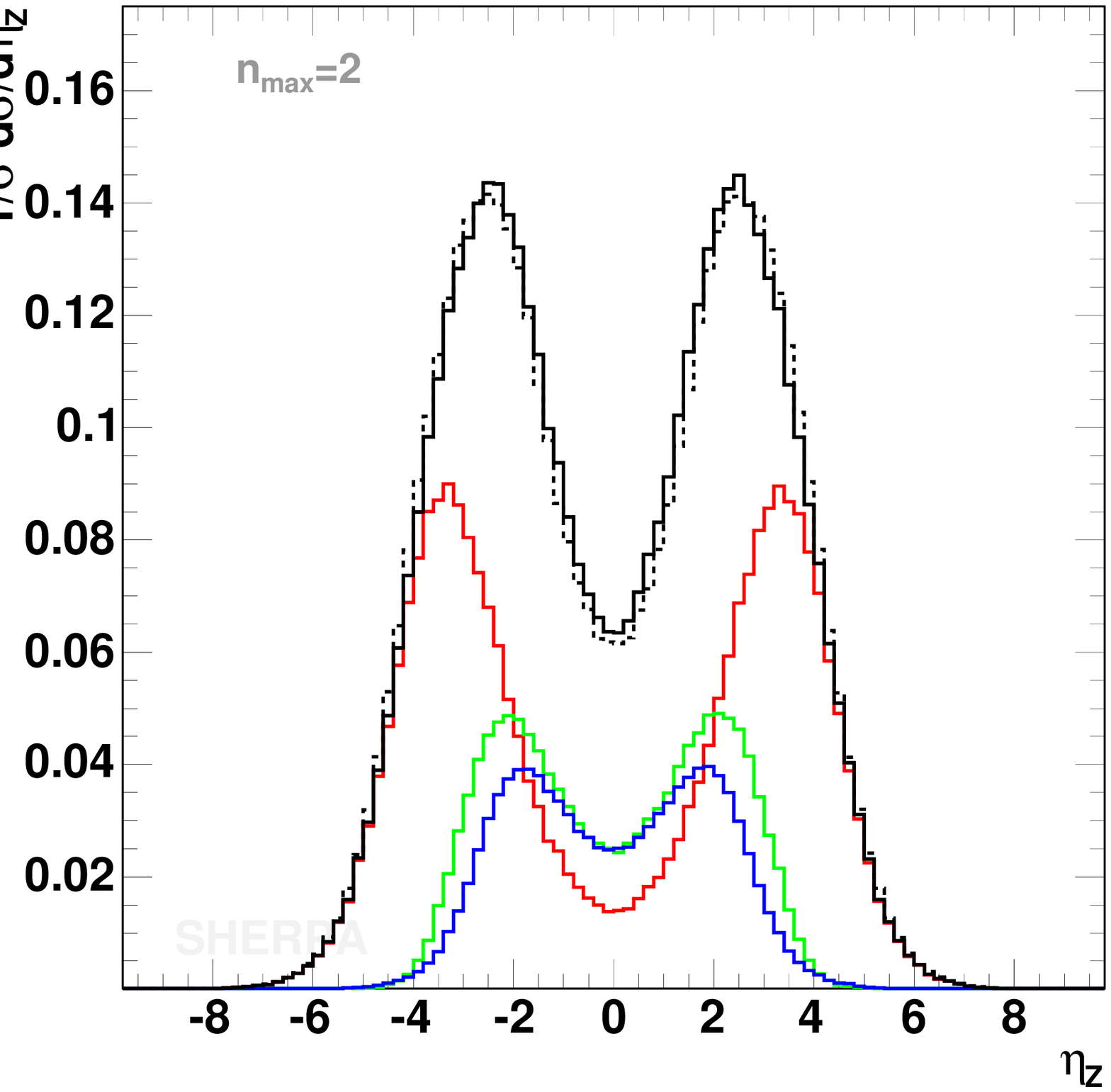}}
\end{pspicture}
\end{center}
\caption{\label{nmax_pt}$p_\perp(Z)$ (upper row) and $\eta(Z)$ (lower row) for 
  $Q_{\rm cut}=15$ GeV and different maximal numbers (2-4, from left to right) 
  of ME jets included. The dashed line corresponds to the maximal 
  number of ME jets reduced by one.}
\end{figure*}
  
\noindent
The effect of varying $n_{\rm max}$ on the transverse momentum and pseudo-rapidity spectrum of 
the lepton pair is exhibited in Fig.\ \ref{nmax_pt}. In this figure, results are compared for 
$n_{\rm max}=2,\,3,\,4$. In each plot, a reference result is given with the corresponding 
$n_{\rm max}^{\rm ref} = n_{\rm max}-1$. For the case of the $p_\perp$ distribution it has already 
been observed that the high-$p_\perp$ region is described through higher multiplicity matrix 
elements. As a consequence, it is the high-$p_\perp$ region that is affected by the variation 
of $n_{\rm max}$. However, while the effect is clearly noticeable when going from 
one to two extra partons the change becomes smaller the more matrix elements are included. From 
the very right plot one can conclude that considering $Z+3$ extra parton matrix elements is a 
reasonable choice to simulate inclusive $Z$ production. The change in the $\eta$ distribution for 
different $n_{\rm max}$ is as expected, considering what has already been seen for varying the jet 
resolution. The higher multiplicity matrix elements favour the region of small $|\eta|$ yielding 
slightly tighter pseudo-rapidity distributions. Again, the more matrix elements have been taken into 
account the smaller the influence when adding an even higher multiplicity. 

\noindent
In comparison to what has been observed when studying gauge boson production at the Fermilab 
Tevatron \cite{Krauss:2004bs}, the LHC provides much more phase space for additional hard QCD 
radiation, enhancing the influence of higher order matrix elements. Therefore a modest value 
of the jet resolution parameter and the inclusion of a sufficient large number of matrix element 
legs is advisable for LHC analyses.
\subsection{Jet observables}

\begin{figure*}[h]
\begin{center}
\begin{pspicture}(400,400)
\put(250,250){\includegraphics[width=5.5cm]{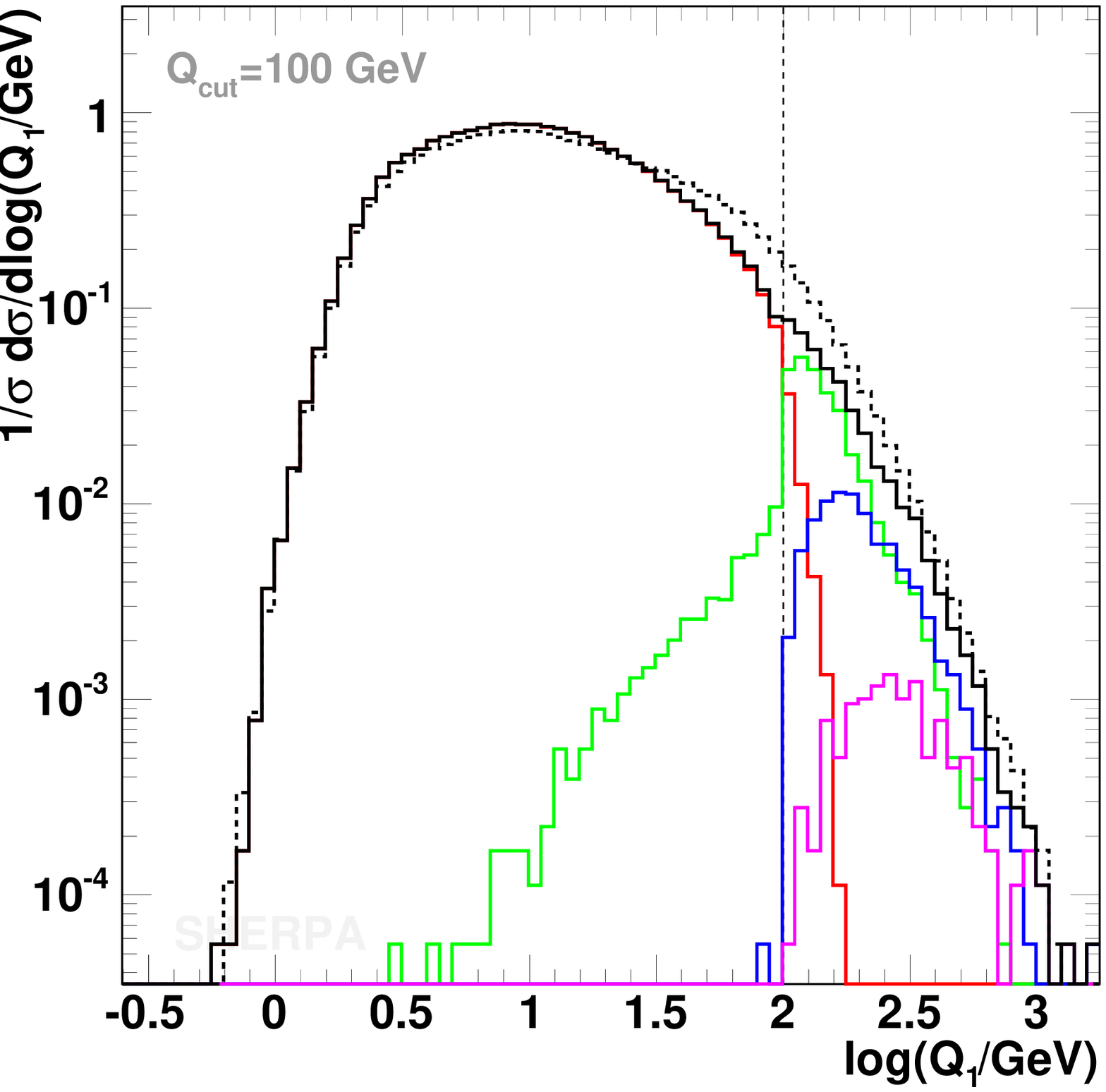}}
\put(125,250){\includegraphics[width=5.5cm]{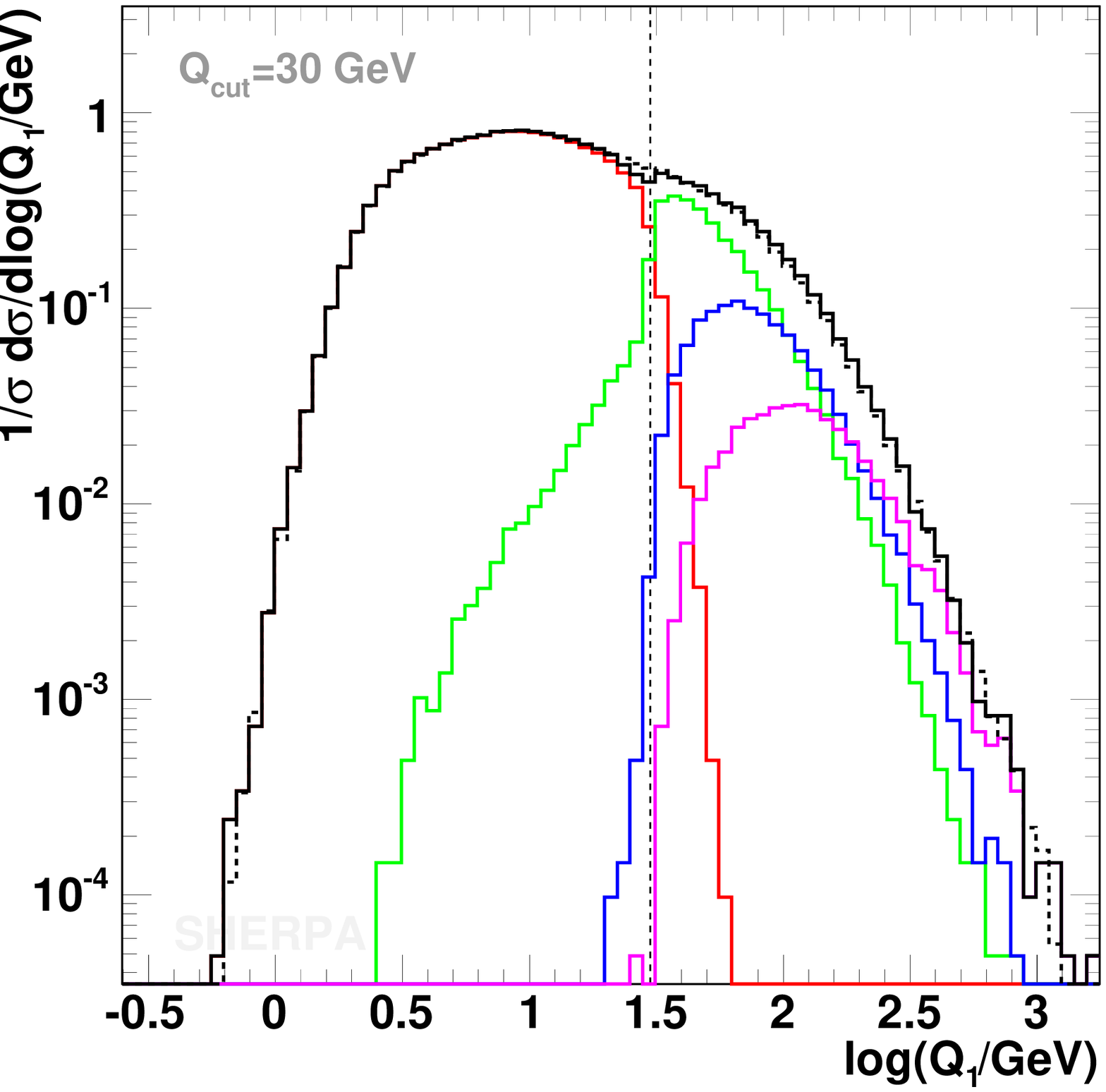}}
\put(0,250){\includegraphics[width=5.5cm]{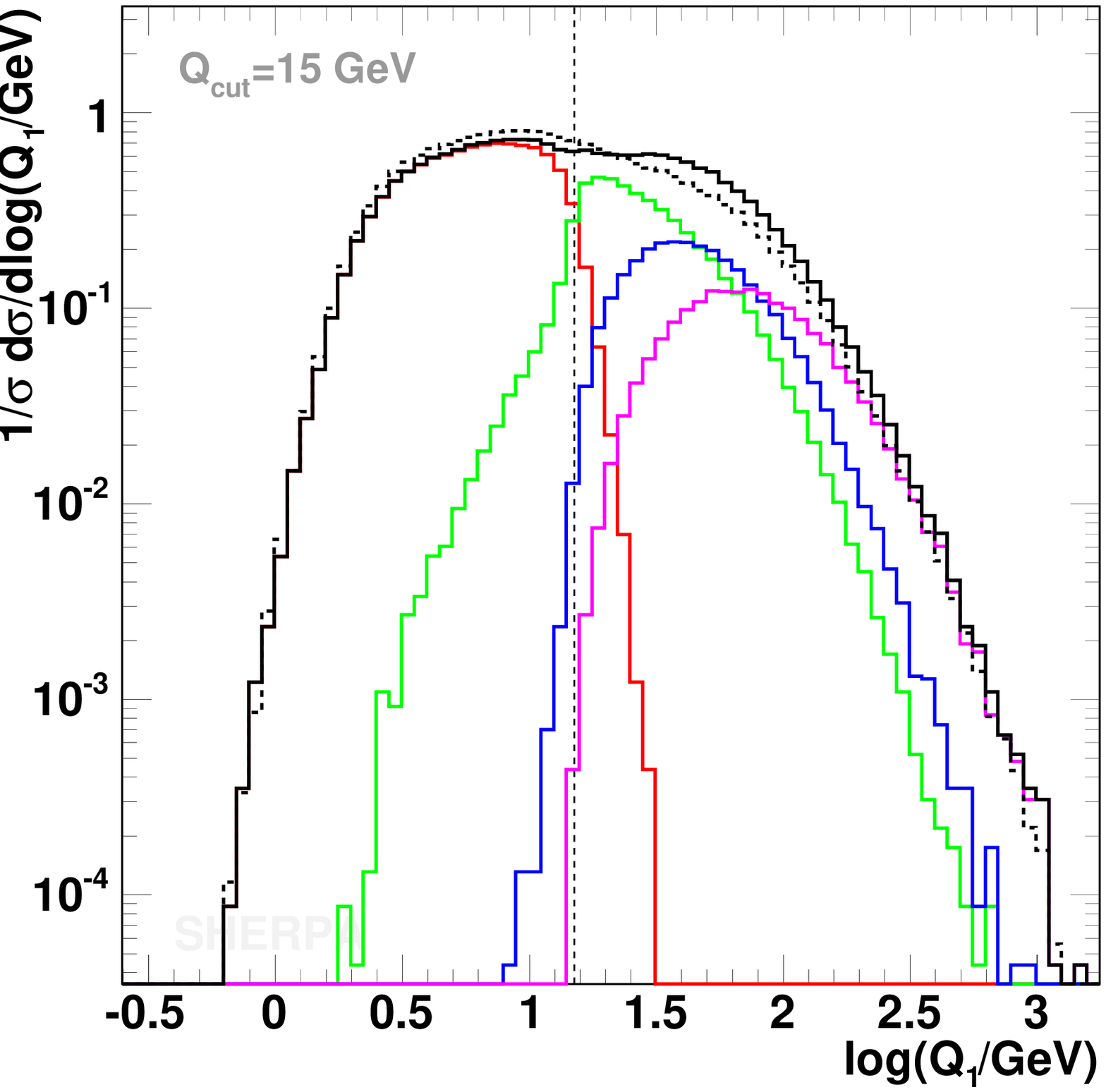}}
\put(250,125){\includegraphics[width=5.5cm]{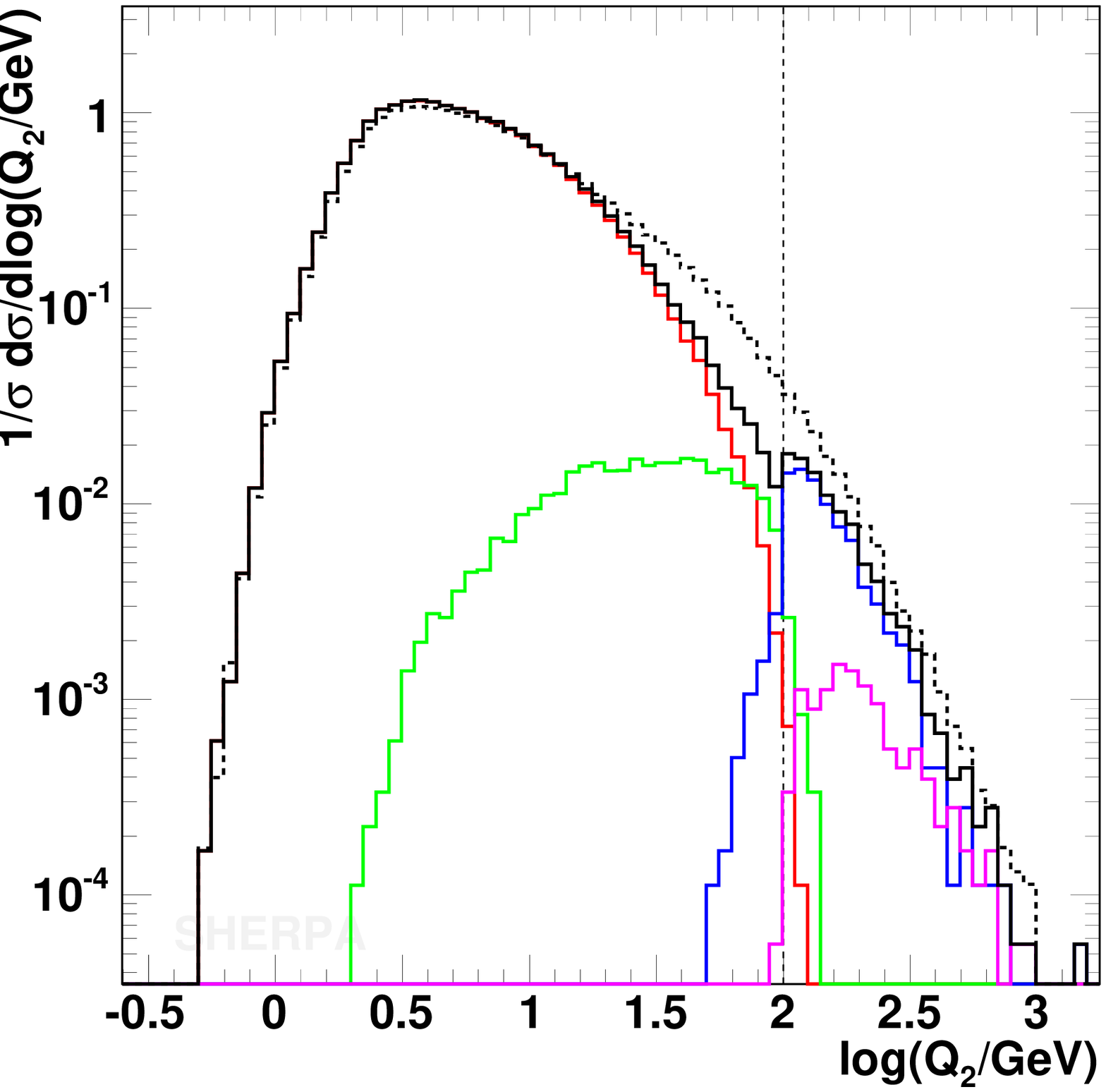}}
\put(125,125){\includegraphics[width=5.5cm]{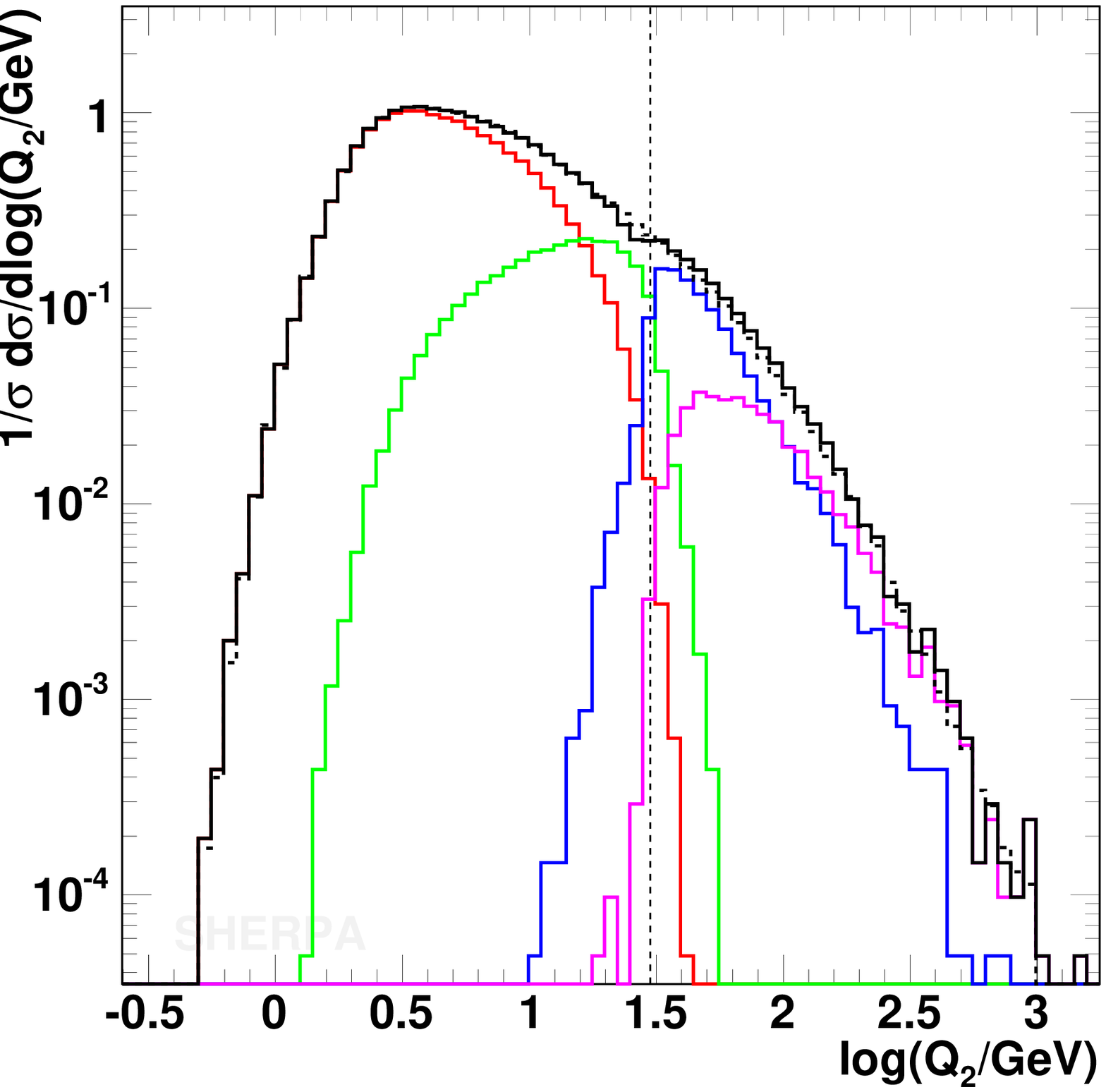}}
\put(0,125){\includegraphics[width=5.5cm]{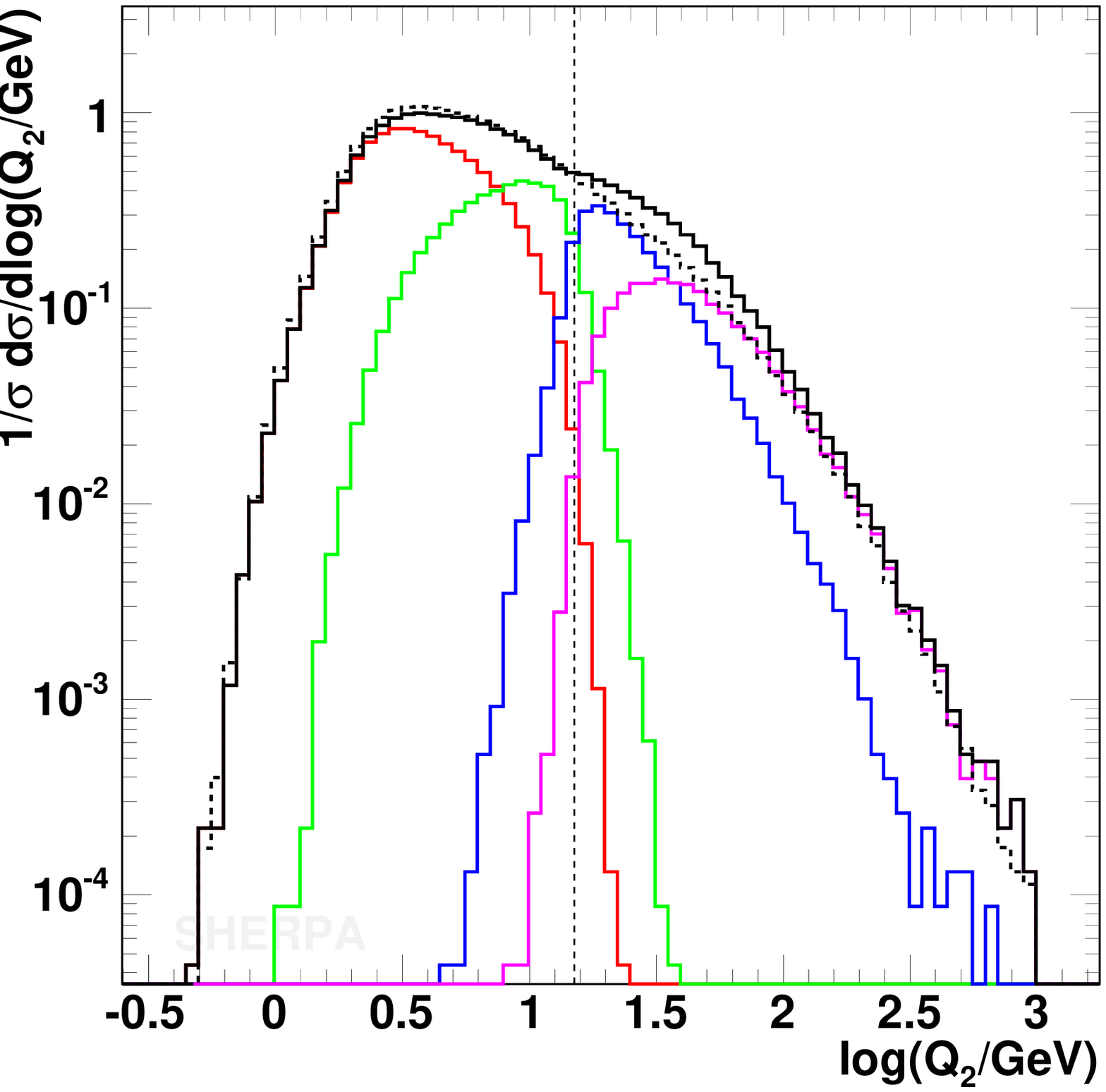}}
\put(250,0){\includegraphics[width=5.5cm]{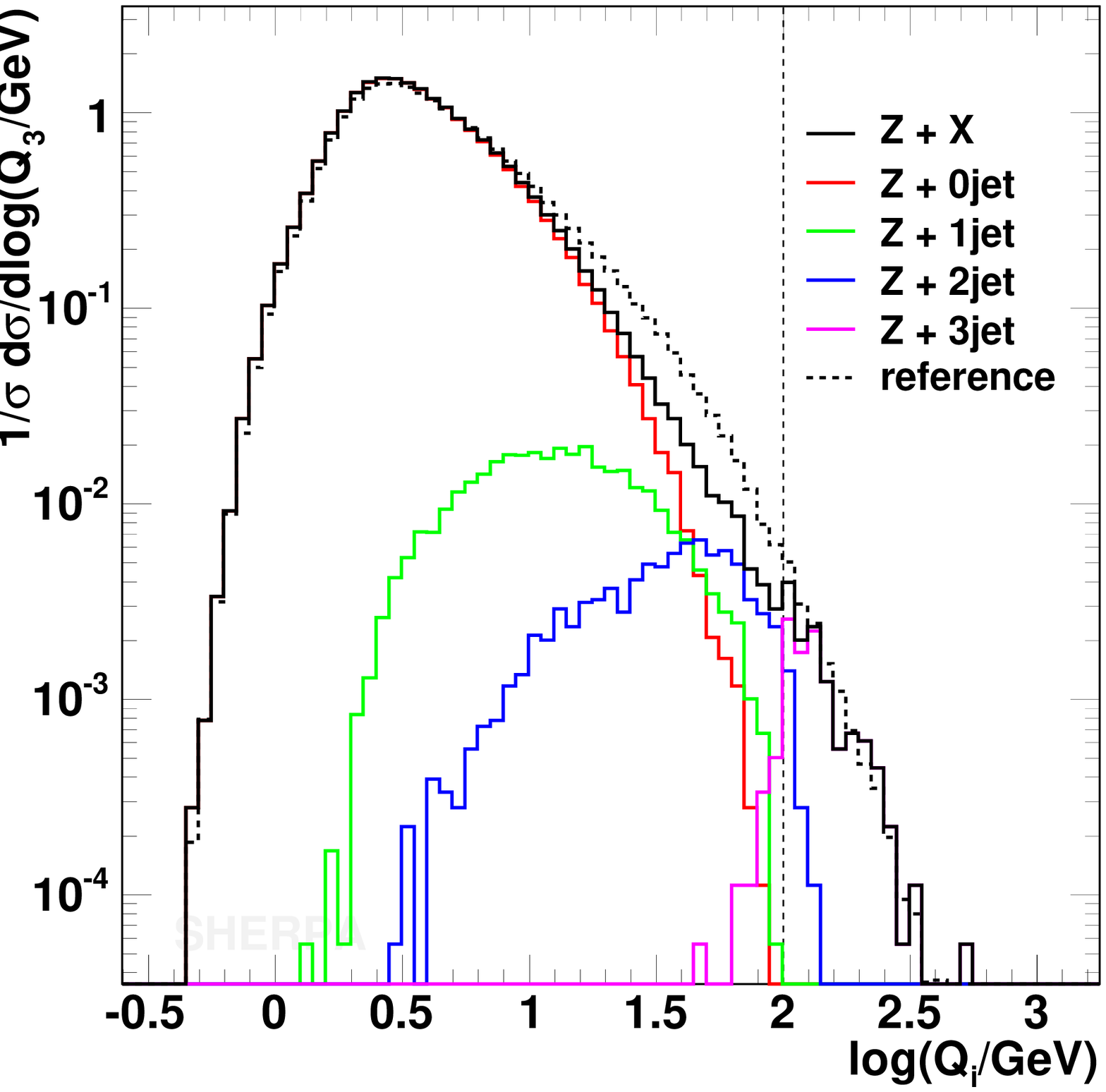}}
\put(125,0){\includegraphics[width=5.5cm]{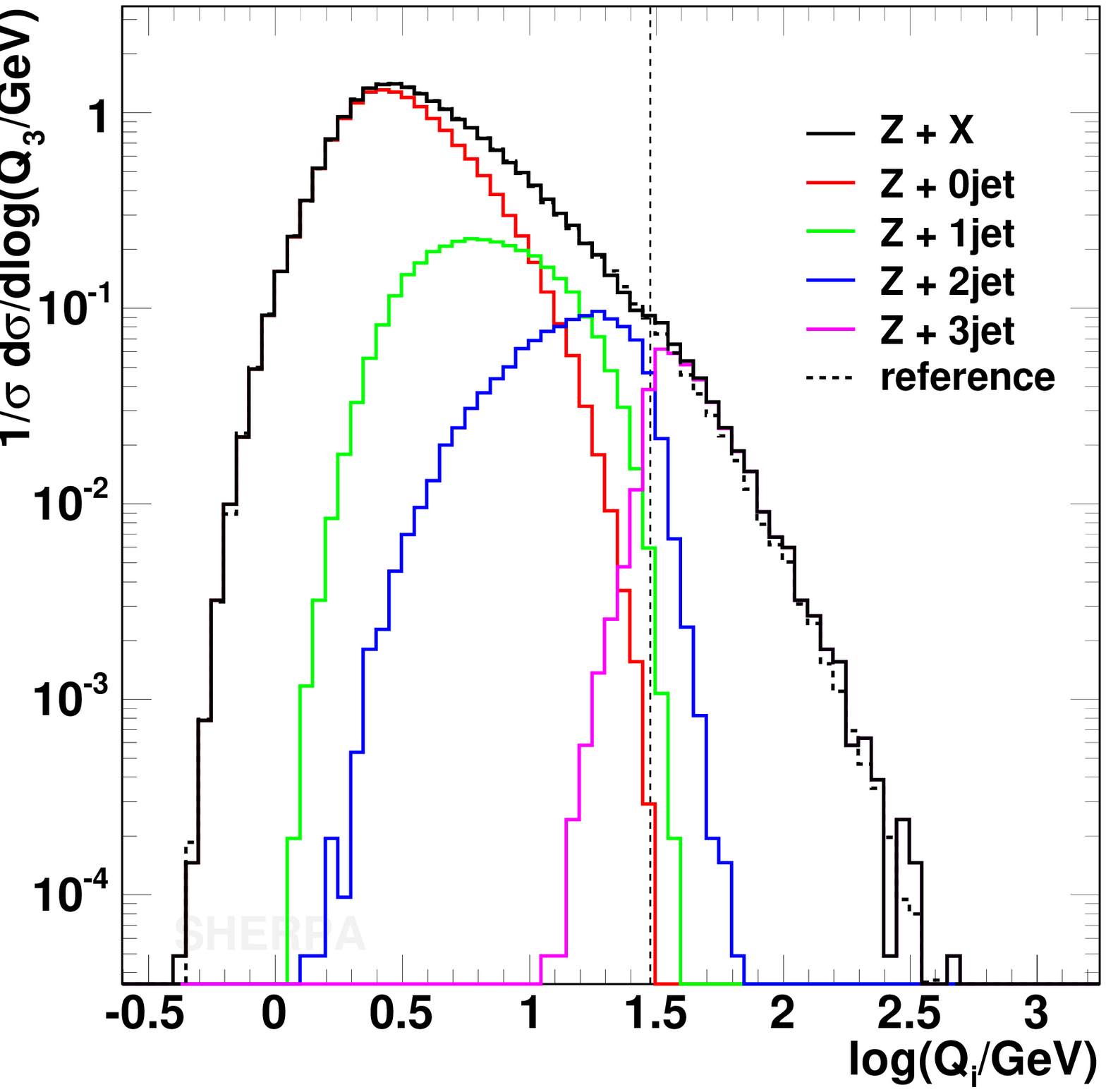}}
\put(0,0){\includegraphics[width=5.5cm]{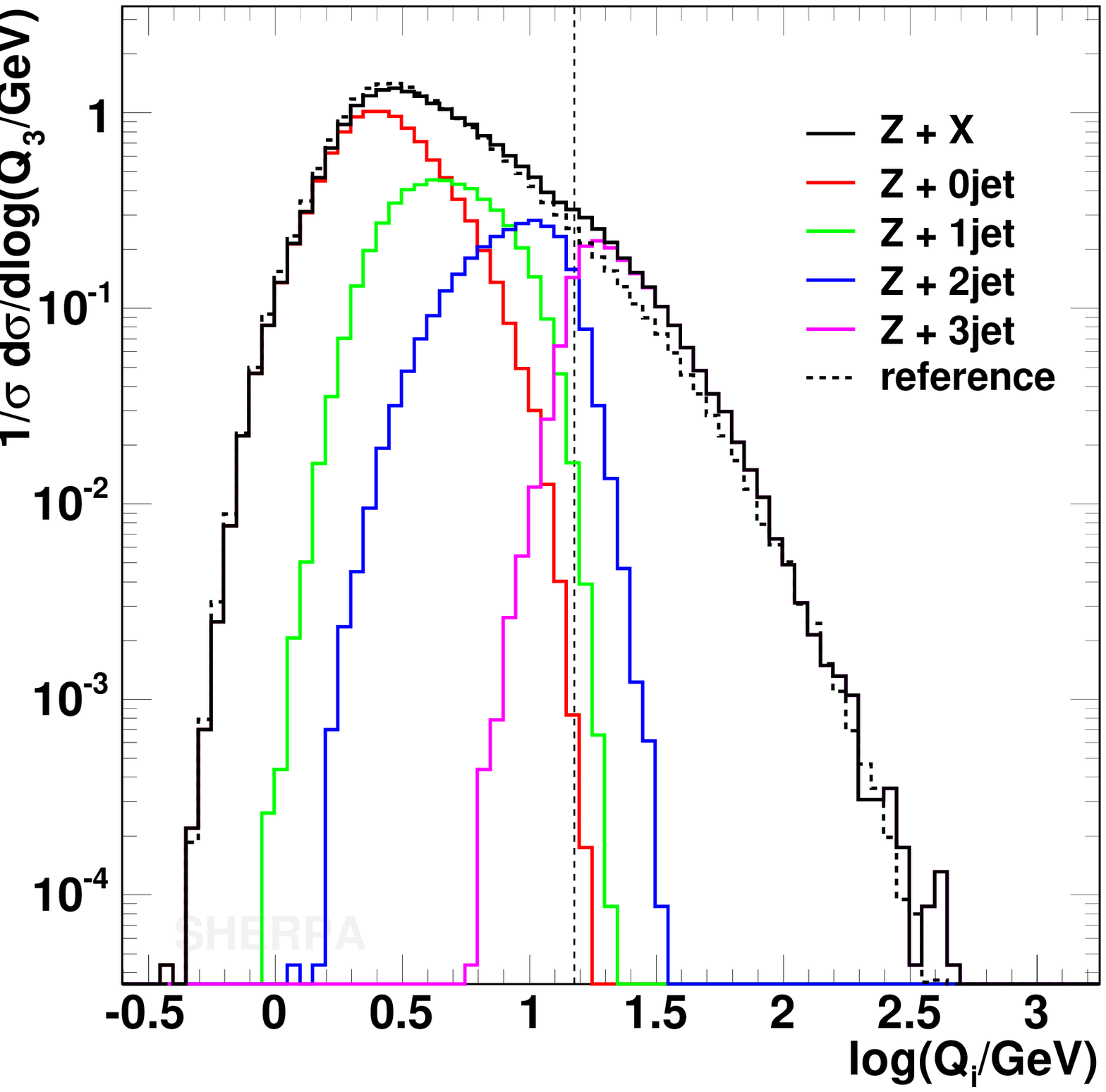}}
\end{pspicture}
\end{center}
\caption{\label{ycut_diff}Differential jet rates for the  $1 \to 0$, 
         $2 \to 1$ and $3 \to 2$ transition (top to bottom), for $Q_{\rm cut}=15$~GeV, 
         $30$~GeV, and $100$~GeV (from left to right). The dashed reference curve in each plot
         is obtained after averaging the corresponding results for 
         $Q_{\rm cut}=15,\,20,\,30,\,50,\,100$ GeV.}
\end{figure*}

\noindent
As has already been seen in the previous publication \cite{Krauss:2004bs}, a very sensitive test 
of the merging procedure is provided by observables based on jets. In particular, differential 
jet rates have turned out to be very useful, since they clearly show how the matrix 
elements and the parton showers interact in filling the phase space below and above the jet 
resolution cut. In Fig.\ \ref{ycut_diff}, differential jet rates using the Run II $k_\perp$ 
clustering algorithm with $R=1$ are depicted. They signal the relevant $Q$ value of the 
$k_\perp$-algorithm, where an $(n+1)$-jet event turns into an $n$-jet event. Again, the results 
for three different values of $Q_{\rm cut}$ are depicted: from left to right, in the columns 
$Q_{\rm cut} = 15,\,30,\,100$ GeV, as indicated by the thin vertical lines. In each plot, 
the resulting spectrum is compared to the average of the results for 
$Q_{\rm cut} = 15,\,20,\,30,\,50,\,100$ GeV. In the three rows, the differential jet rates for 
the $1\to 0$, the $2\to 1$, and for the $3\to 2$ transition (from top to bottom) are shown. 
Starting the discussion with the results for $Q_{\rm cut}=30$ GeV, very good agreement with 
the reference curves can be observed. While the $3\to2$ transition is very smooth around 
the cut the results for $1 \to 0$ and $2 \to 1$ exhibit small dips at the cut scale. Since the 
kinematics of the matrix elements is altered when the parton shower is attached, mismatches of 
the parton configurations close to the cut occur, leading to the dips. Similar structures can be 
observed for the case of $Q_{\rm cut}=100$ GeV. However, more obvious here is that the parton shower 
fails to fill the phase space for hard emissions up to this very large cut. For $Q_{\rm cut}=15$ GeV 
no visible dips at the cut scale are observed. Instead, this sample seems to slightly overestimate 
the contributions from higher order matrix elements w.r.t the reference. A small kink at 
$Q_{\rm cut}$ can be observed for the $1\to 0$ and $2 \to 1$ transition. These residual 
dependences of the results on $Q_{\rm cut}$ may be used to tune the perturbative part of the 
Monte Carlo event generator. 

\begin{figure*}[!t]
\begin{center}
\begin{pspicture}(400,150)
\put(250,0){\includegraphics[width=5.5cm]{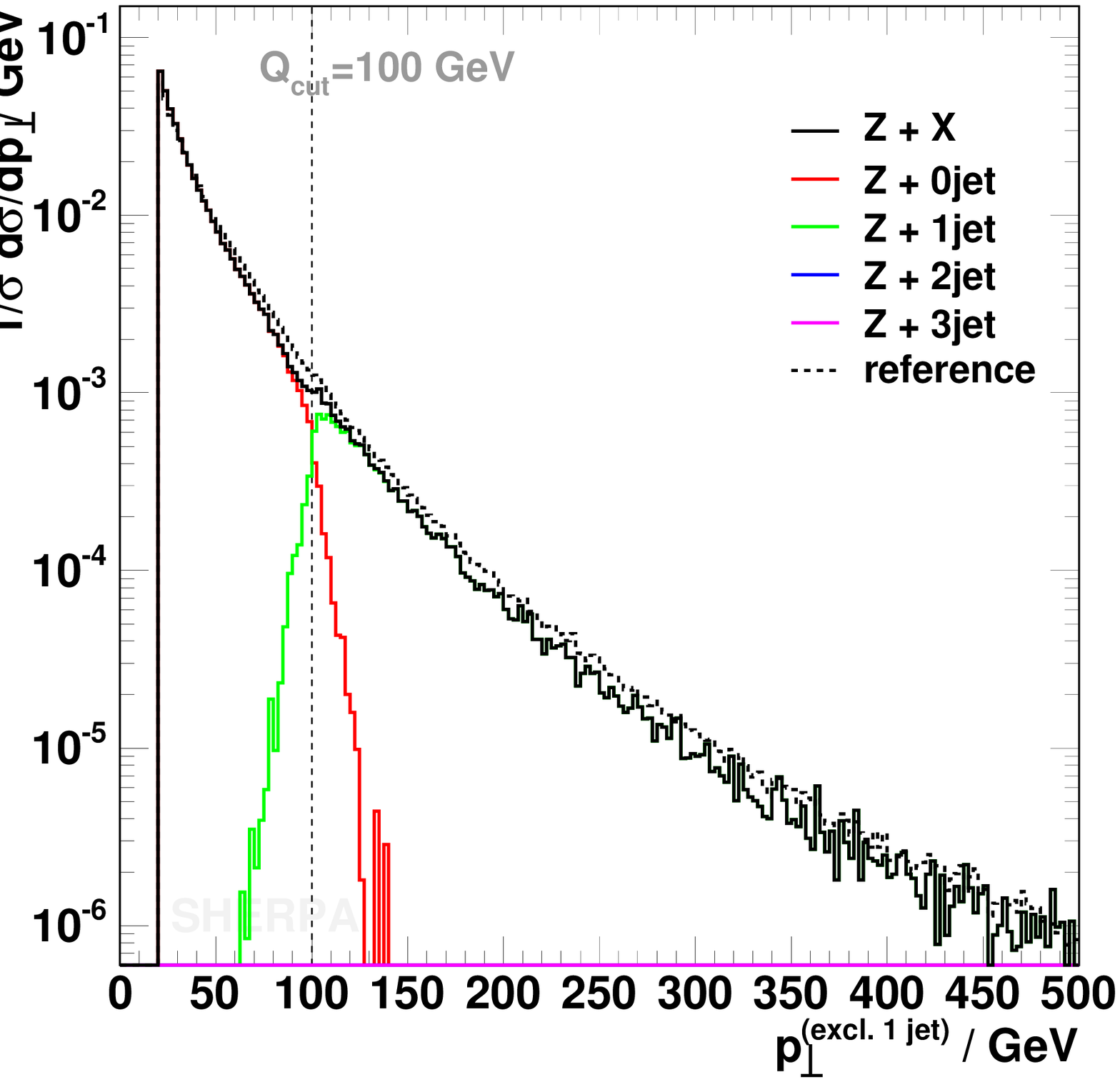}}
\put(125,0){\includegraphics[width=5.5cm]{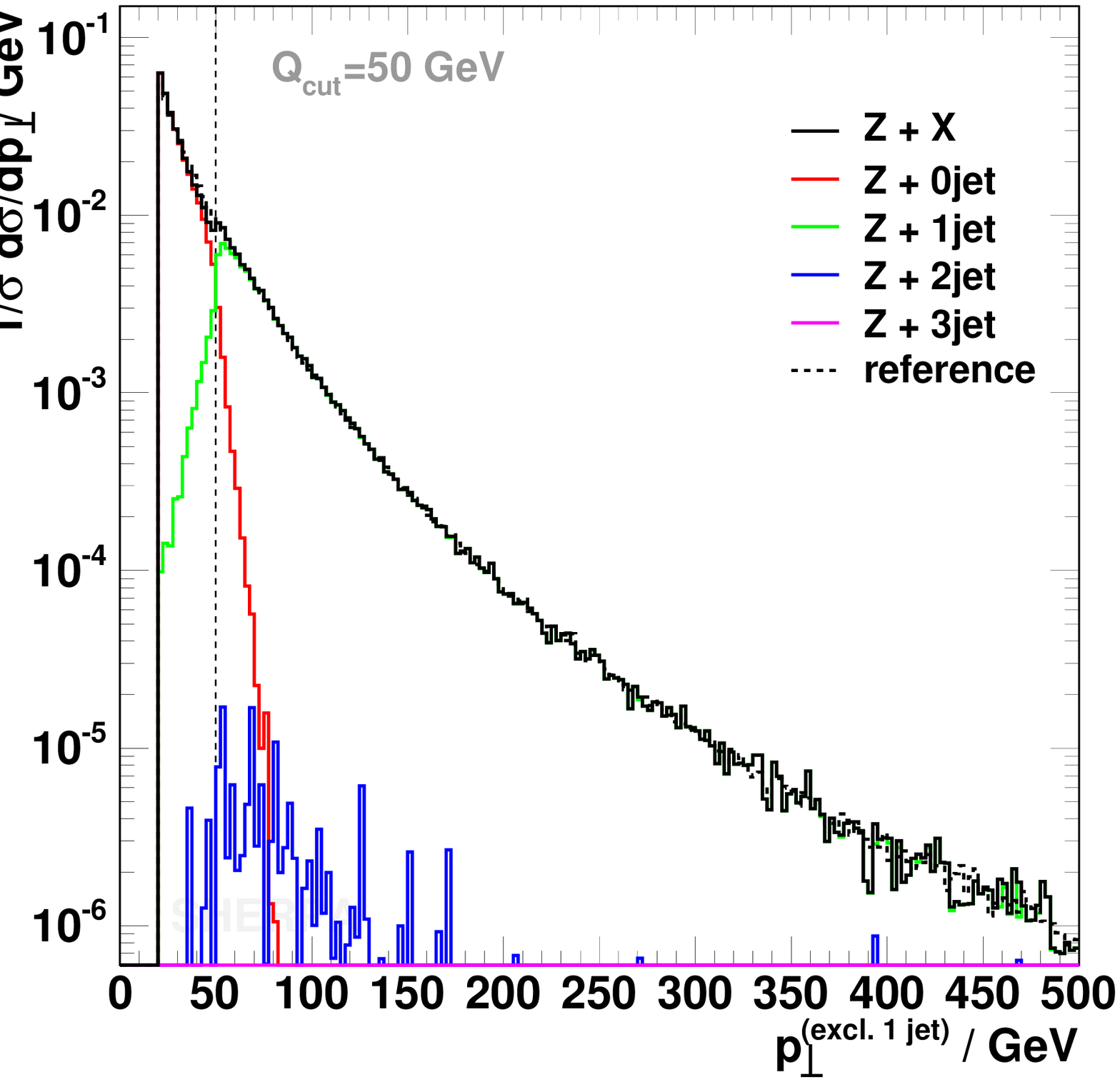}}
\put(0,0){\includegraphics[width=5.5cm]{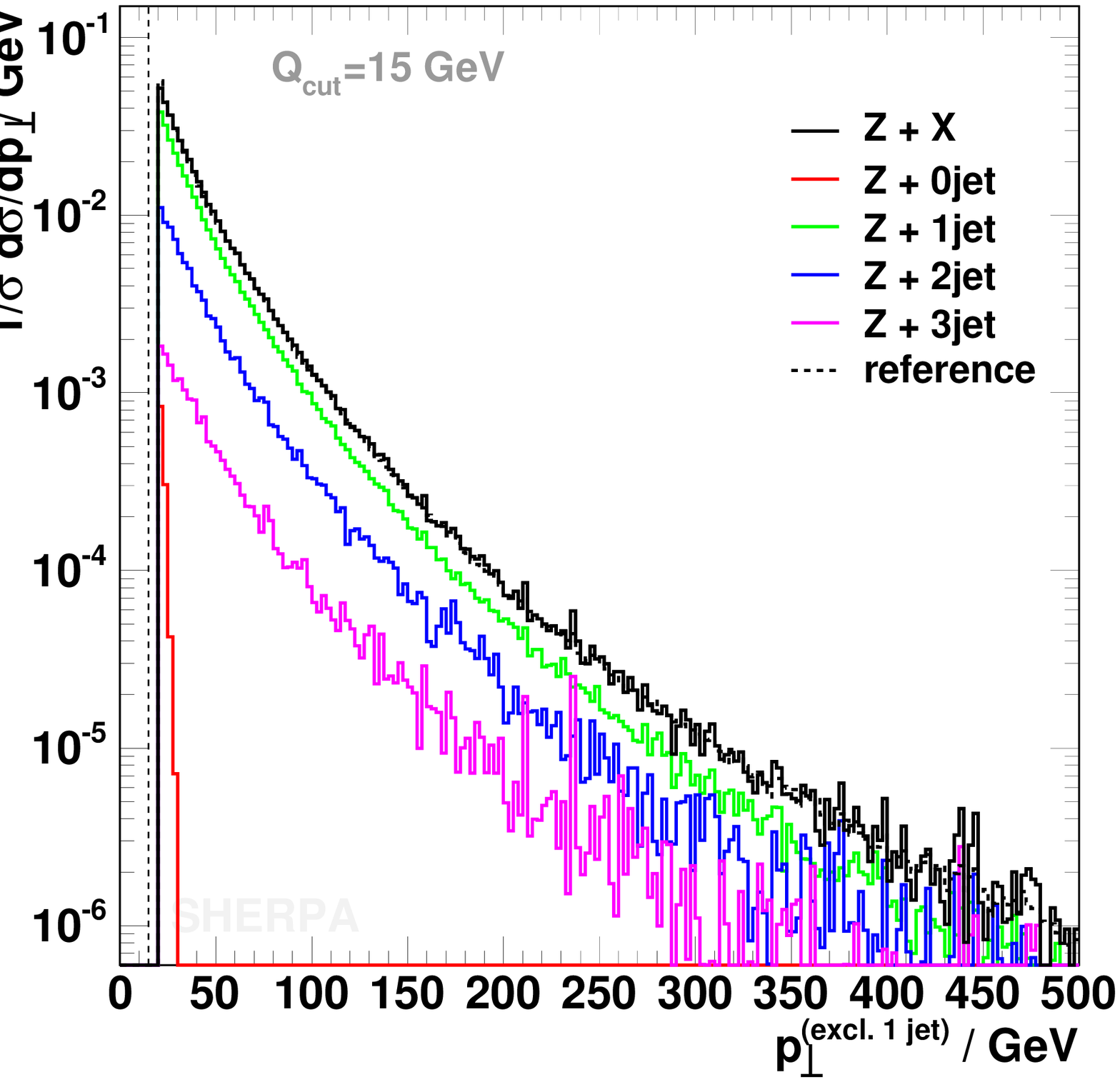}}
\end{pspicture}
\end{center}
\caption{\label{ycut_jets}$p_\perp$ of the jet in exclusive $Z+1$jet production.
  For the jet definition, the Run II $k_\perp$-algorithm with $R=0.4$ and $p_\perp^{\rm jet} > 20$ GeV
  is used. 
  From left to right, results for $Q_{\rm cut}=15,\,50,\,100$ GeV are contrasted with a reference: 
  the average of the results for $Q_{\rm cut} = 15,\,20,\,30,\,50,\,100$ GeV.
  The thin horizontal line indicates the jet resolution scale used.}  
\end{figure*}

\noindent
In Fig.\ \ref{ycut_jets} the $p_\perp$ spectrum of the jet in exclusive $Z+1$jet 
production is shown for three choices of the jet resolution scale, $Q_{\rm cut} = 15,\,50,\,100$ 
GeV, indicated by the thin vertical line. The results are contrasted with a reference 
curve, again the average of results for $Q_{\rm cut} = 15,\,20,\,30,\,50,\,100$ GeV. The 
jet has been defined using the Run II $k_\perp$-algorithm with a minimal jet-$p_\perp$ of $20$ GeV 
and $R=0.4$. The smallest value of $Q_{\rm cut}$ presented here, namely $15$ GeV, is smaller 
than the actual jet cut used in the analysis. Accordingly, matrix elements with more than one 
extra leg have a non-vanishing influence on the jet-$p_\perp$ distribution. This changes as 
soon as $Q_{\rm cut}$ becomes larger than $20$ GeV. For $Q_{\rm cut}=50$ GeV and even more for 
$Q_{\rm cut}=100$ GeV the contributions from matrix elements with $n_{\rm max}>1$ are almost 
negligible. There, only a small dip in the $p_\perp$ distribution around the resolution scale 
can be observed. As has been seen in the transverse momentum distribution of the lepton pair, cf.\ Fig.\ 
\ref{ycut_pt}, for $Q_{\rm cut}=100$ GeV, the shower is not able to fill the full phase space 
below the cut properly. However, the overall agreement of the three results is satisfactory, 
keeping in mind the large parameter range used for $Q_{\rm cut}$.

\begin{figure*}[!t]
\begin{center}
\begin{pspicture}(400,150)
\put(250,0){\includegraphics[width=5.5cm]{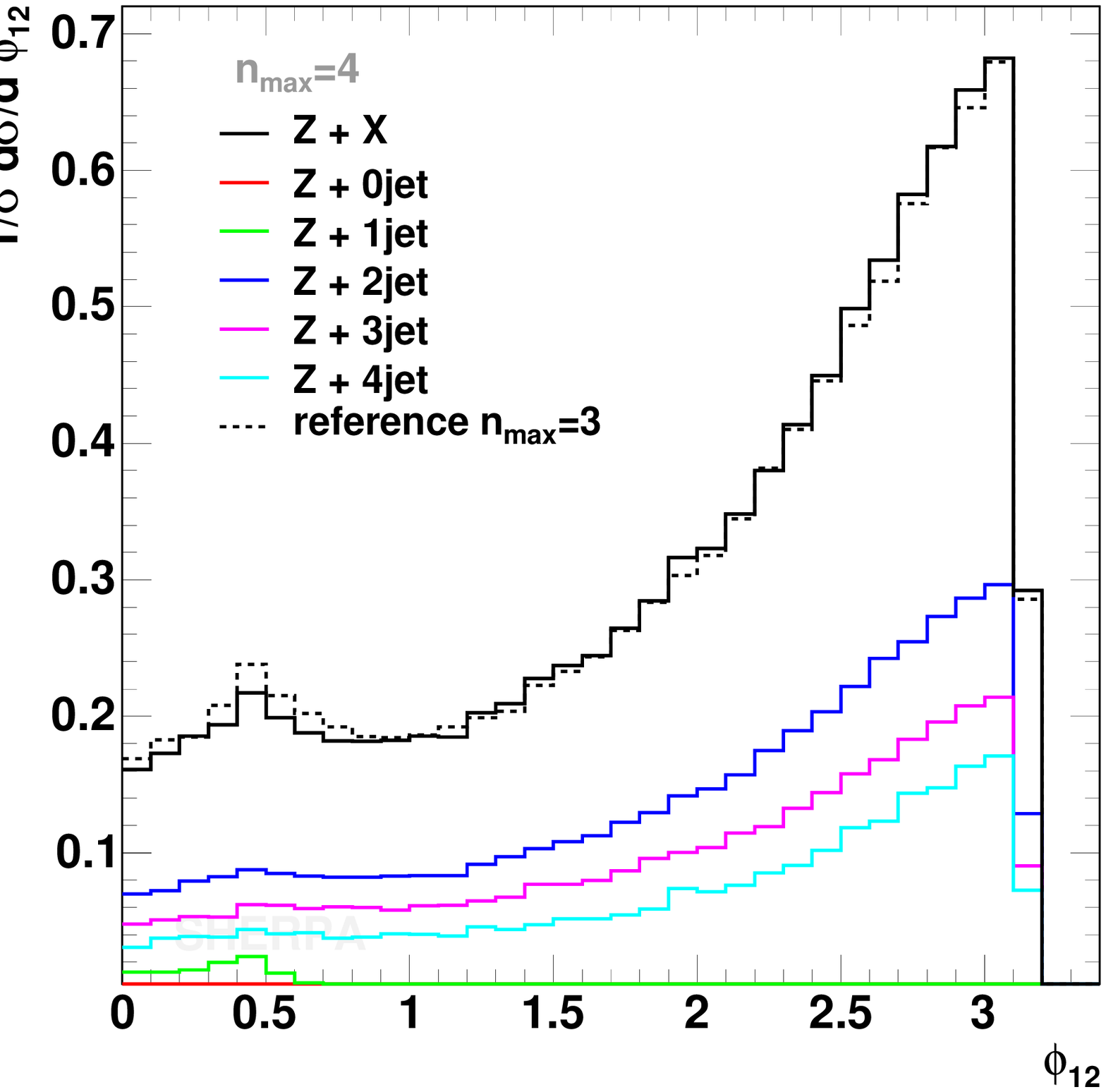}}
\put(125,0){\includegraphics[width=5.5cm]{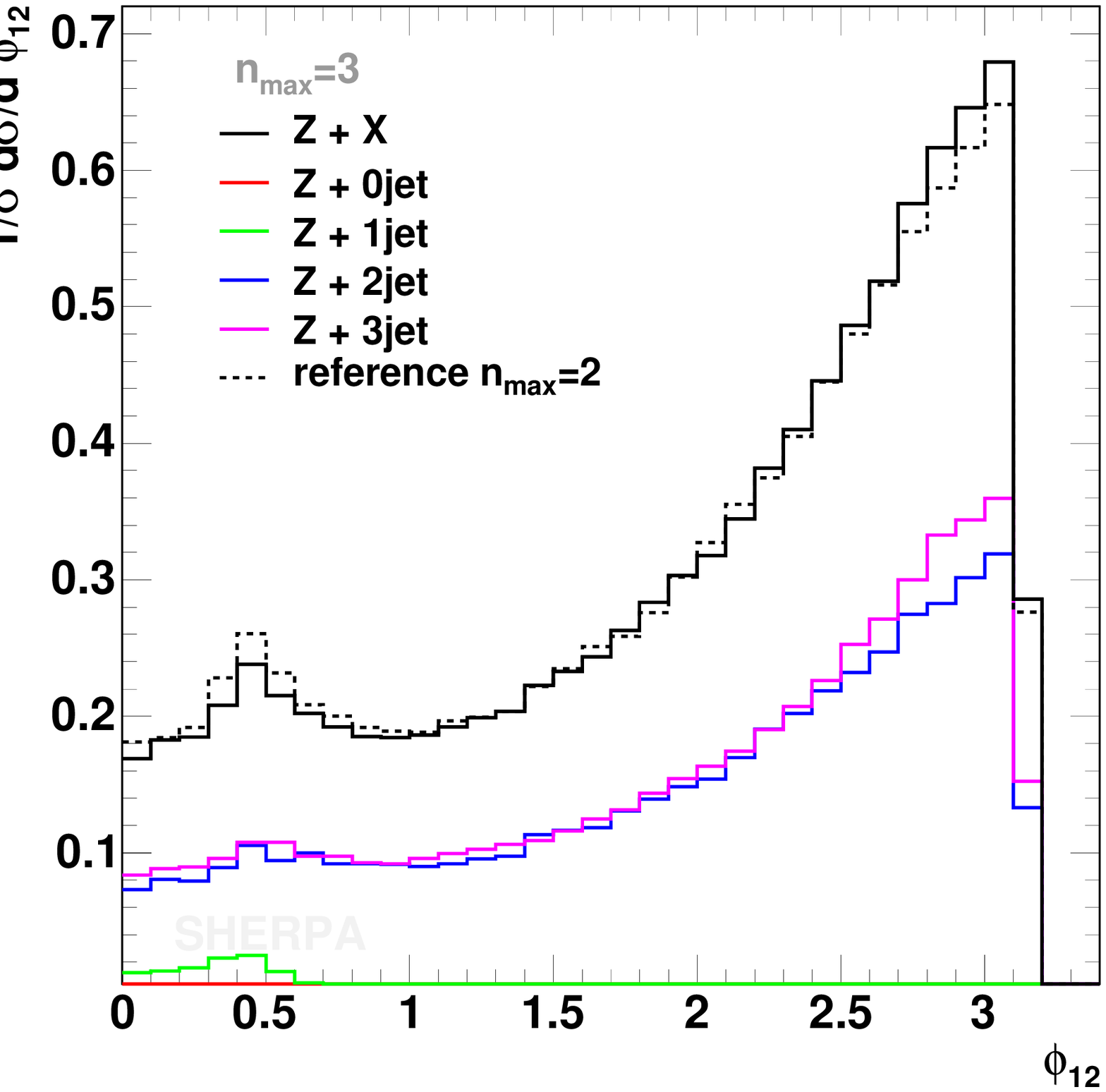}}
\put(0,0){\includegraphics[width=5.5cm]{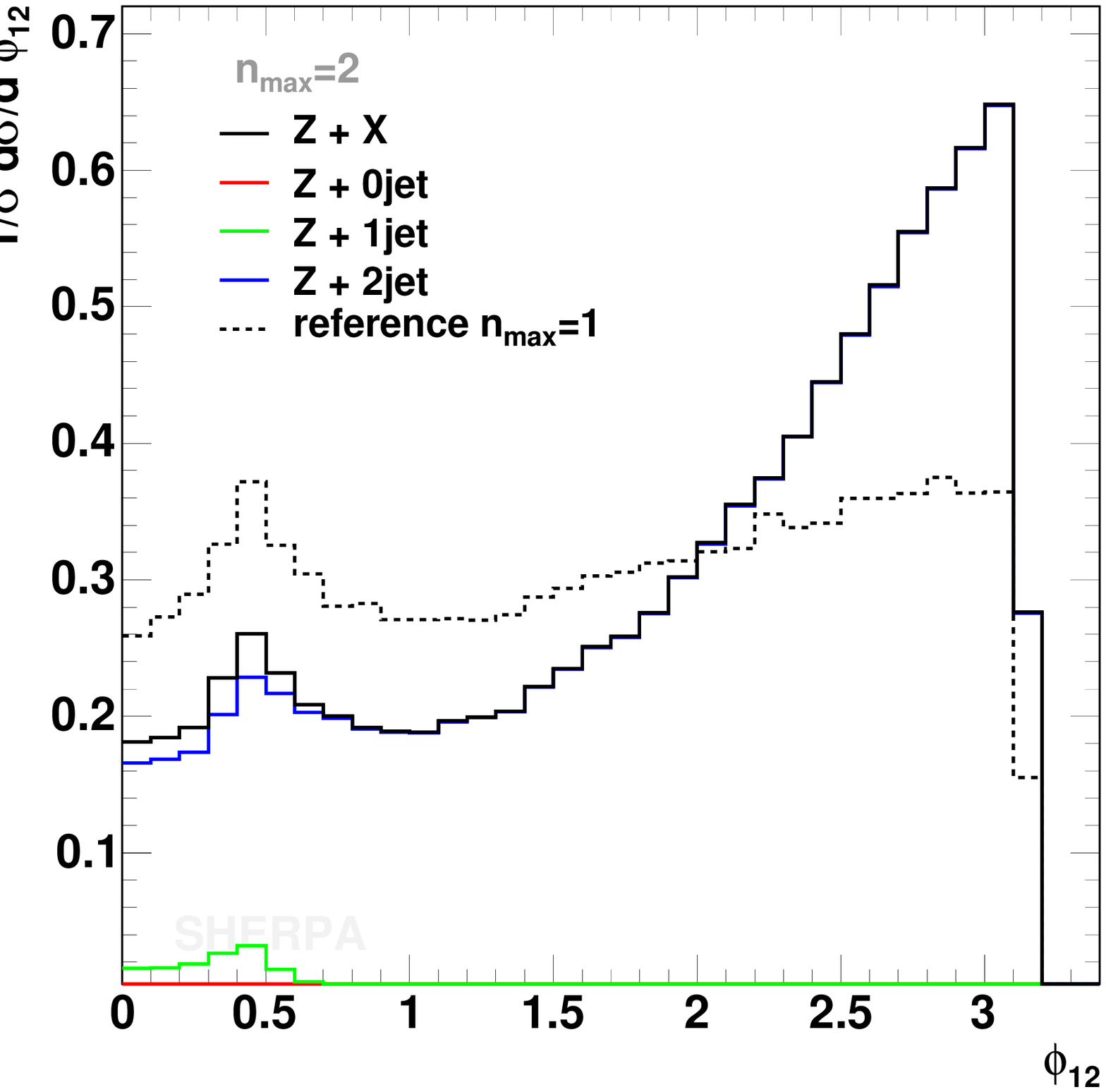}}
\end{pspicture}
\end{center}
\caption{\label{nmax_dphi}$\Delta\phi_{12}$ for $Q_{\rm cut}=15$ GeV and different maximal 
  numbers of ME jets included. The dashed line corresponds to the reference result obtained with
  $n_{\rm max}^{\rm ref} = n_{\rm max}-1$.}
\end{figure*}

\noindent
To highlight the effect of taking into account different maximal numbers of final state 
partons through matrix elements, a two-jet correlation is exhibited in Fig.\ \ref{nmax_dphi}. 
There, the relative transverse angle $\Delta\phi$ between the two hardest jets in 
inclusive $Z+2$jet production is displayed; from left to right, $n_{\rm max}$ has set to 
$n_{\rm max} = 2,\,3,\,4$. Each result is contrasted with a reference that has been obtained 
with $n_{\rm max}^{\rm ref} = n_{\rm max}-1$. From the very left plot it is clear, that the one-jet 
matrix element is incapable of correctly describing the $\Delta\phi$ distribution since the parton 
shower does not treat interferences properly. On the other hand, as soon as $n_{\rm max} \ge 2$, 
the two-jet correlations are consistently described and changes due to the inclusion of higher 
order matrix elements are rather modest. 

\subsection{Variation of renormalisation and factorisation scale}

\begin{figure*}
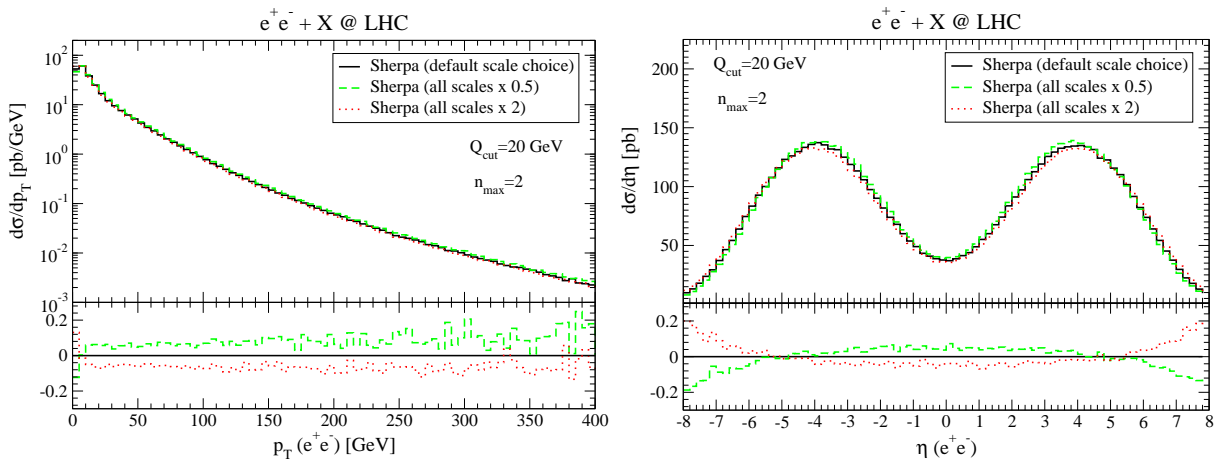

\begin{center}
\begin{tabular}{cc}
\includegraphics[width=8cm]{figures/LHC_Z_pt_scales.eps}&
\includegraphics[width=7.9cm]{figures/LHC_Z_eta_scales.eps}
\end{tabular}
\end{center}
\caption{\label{scale_change1}The transverse momentum (left) and pseudo-rapidity (right) 
distribution of the boson in inclusive $Z/\gamma^*$ production at the LHC and their 
dependence on different choices for the factorisation and renormalisation scale. Results 
have been obtained using $Q_{\rm cut}=20$ GeV and $n_{\rm max}=2$. The black solid lines 
indicate the default hadron level result of \sherpa. To obtain the green dashed (red dotted) 
line, all scales appearing in the coupling constants and the PDFs in both the matrix 
elements and the parton showers are multiplied by a factor of $0.5$ $(2)$. In the lower 
parts of the plots the variations of the results with respect to the default scale choice 
are presented.}
\end{figure*}

\begin{figure*}
\begin{center}
\begin{tabular}{c}
\includegraphics[width=8cm]{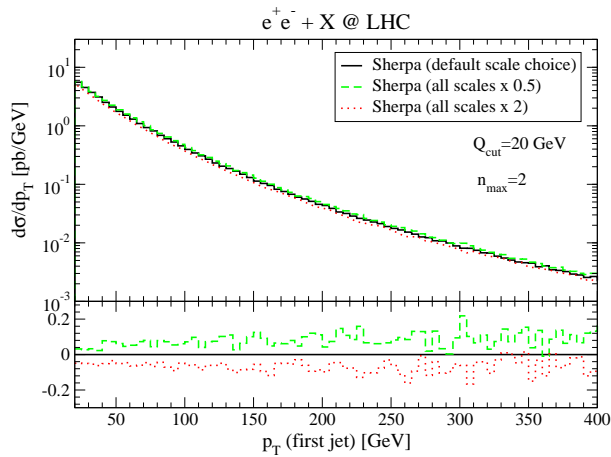}
\end{tabular}
\end{center}
\caption{\label{scale_change2}The $p_\perp$ spectrum of the hardest jet in 
inclusive $Z/\gamma^*$ production at the LHC and its dependence on different choices 
for the factorisation and renormalisation scale. Results have been obtained using 
$Q_{\rm cut}=20$ GeV and $n_{\rm max}=2$. Jets are defined through the Run II 
$k_\perp$-algorithm with $R=0.4$ and $p_\perp^{\rm jet} > 20$ GeV. The black solid 
line corresponds to the default hadron level result of \sherpa. To obtain the green 
dashed (red dotted) line, all scales appearing in the coupling constants and the PDFs 
in both the matrix elements and the parton showers are multiplied by a factor of 
$0.5$ $(2)$. In the lower part of the plot the variations of the results with 
respect to the default scale choice are presented.}
\end{figure*}

\noindent
The algorithm as implemented in \sherpa\ determines the renormalisation and factorisation
scales used in a specific calculation. Of course there is some intrinsic freedom in defining 
the scales used for the evaluation of the PDF or the strong coupling constant. In particular, the 
scales used can be multiplied by constant factors, as long as this alteration is applied both
in the matrix element evaluation and reweighting and in the parton shower. This restriction
is due to the construction of how the leading logarithmic dependence on $Q_{\rm cut}$ is 
eliminated. The dependence of the \sherpa\ results with respect to such scale variations is studied
 in Fig.\ \ref{scale_change1} and Fig.\ \ref{scale_change2}. Results obtained with the default 
scale choices are confronted with results obtained when all scales appearing in the coupling 
constants and PDFs are multiplied by common factors of $0.5$ and $2$. 

\noindent
In Fig.\ \ref{scale_change1} the transverse momentum and pseudo-rapidity distribution of the 
$Z/\gamma^*$ boson are depicted. For the case of the $p_\perp$ spectrum, except for the very 
first bins the spectrum obtained with a factor of $0.5$ ($2$) is always above (below) the 
default result. The differences are rather constant and of the order of $10 - 15\%$. As has 
been seen before, cf. Fig.\ \ref{ycut_pt} and \ref{nmax_pt}, for transverse momenta above the 
cut scale the distribution is predominantly described by higher order matrix elements, 
whose scale dependence is known to be tamed with respect to the lowest order process 
\cite{Campbell:2003hd}. This lowest order process, however, dominates the region of very low 
boson momenta. There the $2 \to 2$ cross section exhibits a strong decline when the scales become 
smaller. This effect potentially leads to the reversal of the discrepancies in the soft region. 
The situation in the case of the pseudo-rapidity distribution is very similar. 
From Fig.\ \ref{ycut_eta} and Fig.\ \ref{nmax_pt} one can read off that the region of large 
values of $|\eta|$ is described by the parton shower attached to the $2 \to 2$ matrix element. 
For $|\eta|>5$ the spectrum, where all scales have been multiplied by a factor of two, is 
enhanced up to $20\%$. A factor of $0.5$ on the other hand depopulates 
this phase space region by up to $20\%$. In the intermediate range of pseudo-rapidity the 
deviations of the two spectra from the default scale choice are well below $10\%$.  

\noindent
In Fig.\ \ref{scale_change2} the transverse momentum distribution of the hardest jet
in inclusive $Z/\gamma^*$ production is depicted. In contrast to the two distributions 
above, this result has no significant contribution from the leading order $2 \to 2$ process. 
Therefore, the two results obtained after scale manipulation do not cross each other. 
Over the whole range of jet transverse momentum the deviations of the two curves from the 
default result are very moderate. 

\noindent
It can be concluded that the predictions of \sherpa\ show rather mild variations 
over a wide range of the phase space when multiplying all scales appearing in the coupling 
constants and PDFs by common factors of $0.5$ and $2$. The largest deviations from the default 
choice of scales are observed in those phase space regions that are predominantly
covered by the $2 \to 2$ matrix element with the parton shower attached. 

\section{SHERPA vs.\ NLO results}\label{Compare_NLO_Sec}

\noindent
Having investigated the self-consistency of the merging procedure as implemented in \sherpa\, 
its parton level results are compared with those from {\tt MCFM}, {\tt v.\ 4.0}, 
\cite{Campbell:2002tg,Campbell:2003hd}. For the class of processes studied here, {\tt MCFM} is 
capable to calculate total and fully differential cross sections at next-to-leading order in the 
strong coupling constant for $(Z/\gamma^*\to l^+l^-)+0,1,2$ and $(W^\pm\to l\nu_l)+0,1,2$ partons.
For all calculations with {\tt MCFM} the cteq6l \cite{Pumplin:2002vw} PDF has been used, and $\alpha_S(m_Z)=0.118$ in
accordance with the value of the PDF evolution. The renormalisation and factorisation scales have 
been chosen to be identical with the bosons mass, i.e.\ $\mu_R=\mu_F={\rm m_Z}$ or 
$\mu_R=\mu_F={\rm m_W}$, respectively. Phase space cuts are listed in the appendix \ref{appendix}. 
In contrast to a previous publication \cite{Krauss:2004bs}, this time only ``inclusive'' 
quantities are compared. For the next-to-leading order calculation this translates 
into an unconstrained phase space for the real higher order correction. Thus, the higher order 
corrections may give rise to an additional jet. The \sherpa\ results were obtained after the 
parton shower evolution. For the sake of a better comparison, all curves have been normalised 
to one, eliminating the enhancement of the cross section due to the NLO corrections.

\begin{figure*}
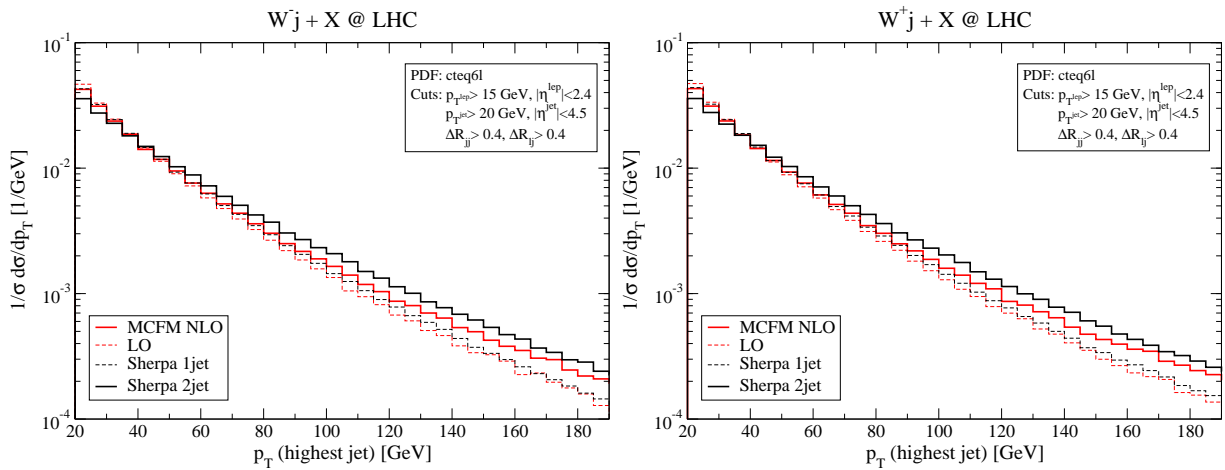

\begin{center}
\begin{tabular}{cc}
\includegraphics[width=8cm]{figures/LHC_Wm1jet_incl.eps}&
\includegraphics[width=8cm]{figures/LHC_Wp1jet_incl.eps}
\end{tabular}
\end{center}
\caption{\label{Wplus1pt}The $p_T$ distribution of the hardest jet for inclusive 
  $W^-$ (left panel) and $W^+$ (right panel) plus one jet events at the LHC.}
\end{figure*}

\begin{figure*}
\begin{center}
\includegraphics[width=8cm]{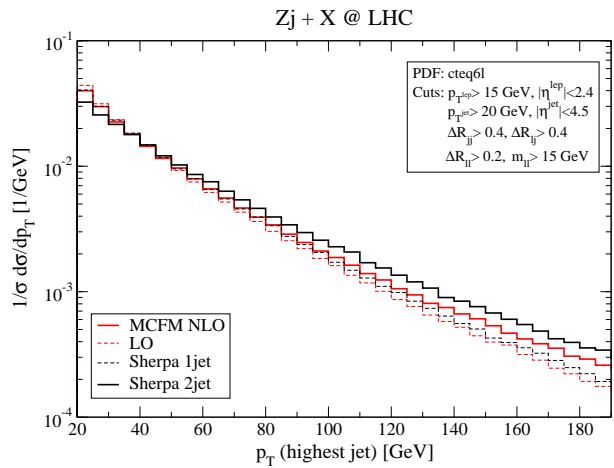}
\end{center}
\caption{\label{Zplus1pt}The $p_T$ distribution of the hardest jet for inclusive 
  $Z/\gamma^*$ plus one jet events at the LHC.}
\end{figure*}

\noindent
First of all, in Figs.\ \ref{Wplus1pt} and \ref{Zplus1pt} the $p_\perp$ spectra of the hardest jet 
in inclusive $W^++1$jet, $W^-+1$jet, and $Z/\gamma^*+1$jet production are exhibited. For all cases, 
results at leading and at next-to-leading order were contrasted with results from \sherpa\ that were 
obtained with $n_{\rm max}=1$ and with $n_{\rm max}=2$, respectively. In all plots the high-$p_\perp$ 
tail is significantly enhanced when going from LO to NLO. The \sherpa\ samples with $n_{\rm max}=2$ 
show the same behavior but tend to pronounce the high-$p_\perp$ region even more. This is in striking 
contrast to the $n_{\rm max}=1$ samples. They are incapable of recovering the shape of the distribution 
at NLO, and tend to look like the LO result. This is not surprising. The NLO calculation takes into 
account tree-level matrix elements with two final state partons as the real contribution to the NLO 
result. Due to the large phase space available at the LHC this real contribution tends to 
produce an extra jet that alters the kinematics of the first jet. Obviously this significant 
change in the kinematics can not be appropriately recovered by the parton shower. The 
$n_{\rm max}=2$ \sherpa\ samples also include the parton shower, resulting in 
increased parton emission thus enhancing the high-$p_\perp$ tail even more. It 
would for sure be instructive to check this behaviour with a resummed NLO 
computation for these processes.

\begin{figure*}
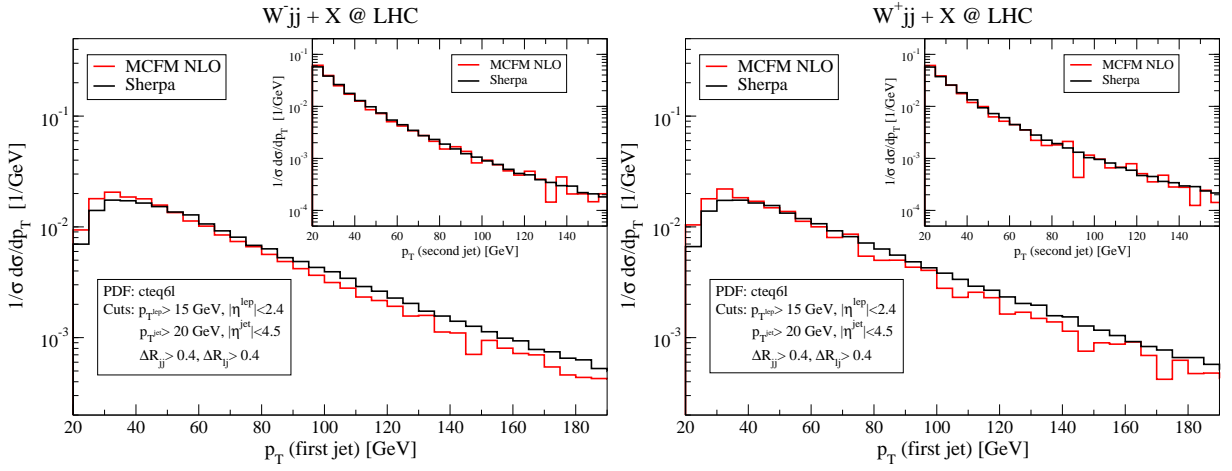

\begin{center}
\begin{tabular}{cc}
\includegraphics[width=8cm]{figures/LHC_Wm2jet_incl.eps}&
\includegraphics[width=8cm]{figures/LHC_Wp2jet_incl.eps}
\end{tabular}
\end{center}
\caption{\label{Wplus2pt}The $p_T$ distribution of the two hardest jets in inclusive 
  $W^-$ (left panel) and $W^+$ (right panel) plus two jet production at the LHC.}
\end{figure*}

\begin{figure*}
\begin{center}
\includegraphics[width=8cm]{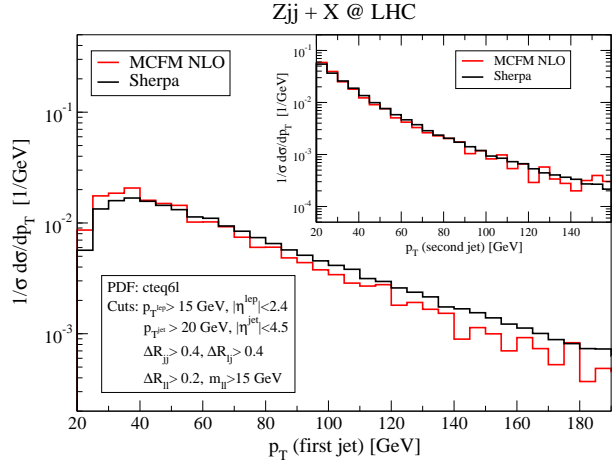}
\end{center}
\caption{\label{Zplus2pt}The $p_T$ distribution of the two hardest jets in inclusive 
  $Z/\gamma^*$ plus two jet production at the LHC.}
\end{figure*}

\begin{figure*}
\begin{center}
\includegraphics[width=8cm]{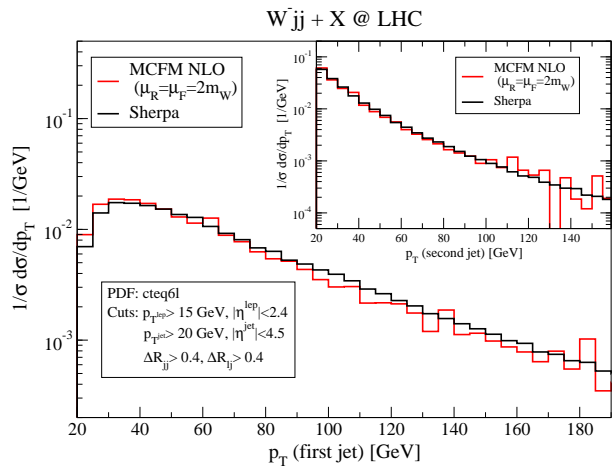}
\end{center}
\caption{\label{Wplus2pt_scale2}The $p_T$ distribution of the first and second hardest jet in 
  inclusive $W^-$ plus two jet production at the LHC. For the NLO calculation the renormalisation 
  and factorisation scale have been chosen to $\mu_R=\mu_F=2\,{\rm m_W}$.}
\end{figure*}

\noindent
In Figs.\ \ref{Wplus2pt} and \ref{Zplus2pt} the $p_\perp$ spectra of the two
hardest jets in inclusive $W^++2$jet, $W^-+2$jet, and $Z/\gamma^*+2$jet production 
are displayed. This time next-to-leading order results from {\tt MCFM} are compared
with the corresponding \sherpa\ samples with $n_{\rm max}=2$. It has been shown
in \cite{Campbell:2003hd} that the shapes of the distributions when going from LO to 
NLO are quite stable. The slopes of the next-to-leading order and the \sherpa\ result 
are in good agreement, the latter having the tendency to produce the first jet 
slightly harder. In Fig.\ \ref{Wplus2pt_scale2} the  $p_\perp$ spectra of the two 
hardest jets in inclusive $W^-+2$jet production are displayed once more. This time, 
however, the renormalisation and factorisation scale in the NLO calculations has 
been chosen as $\mu_R=\mu_F=2\,{\rm m_W}$. For this choice of the scales the agreement 
of {\tt MCFM} and \sherpa\ is even better. This highlights the effect of scale 
variations, a good way to estimate residual uncertainties due to higher order 
corrections and shows that the results of \sherpa\ are well within theoretical 
uncertainties {\footnote {It should be noted that the effect of this scale variation 
on the total cross section merely is of the order of $1\%$, although the shape of 
the distribution in the high-$p_\perp$ tail changes considerably.}}. 

\section{SHERPA vs.\ MC@NLO and PYTHIA}\label{Compare_MC_Sec}

\noindent
In this section, hadron-level results of \sherpa\ will be compared with those of two other 
event generators, namely {\tt MC@NLO} \cite{Frixione:2002ik,Frixione:2003ei,Frixione:2004wy} 
and {\tt PYTHIA} \cite{Sjostrand:1993yb,Sjostrand:2001yu}. The former program incorporates a
consistent matching of a full-fledged next-to-leading order calculation with the parton
shower provided by {\tt HERWIG} \cite{Corcella:2000bw,Corcella:2002jc}. It thus employs 
an angular-ordered shower, taking full account of coherence effects. In contrast, 
{\tt PYTHIA} uses tree-level matrix elements, in this case for $q\bar q\to e^+e^-$ and it 
employs a virtuality-ordered parton shower to model further emissions. In this framework, 
coherence effects are approximated through an explicit veto on rising opening angles in 
the splitting. Hence, the parton shower implementations of {\tt PYTHIA} and \sherpa\ 
are quite similar. However, in order to account in {\tt PYTHIA} for jets with a $p_\perp$ 
larger than the ``natural'' starting scale of the parton shower equal to the invariant 
mass of the lepton pair, the starting scale has been increased to the centre-of-mass 
energy of the  proton-proton system, i.e.\ to 14 TeV. This choice is supplemented with 
a matrix element correction procedure implemented through reweighting meant to reproduce the 
exact matrix element for the emission of an additional jet. The precise setups for both codes 
can be found in appendix \ref{appendix_mc}.

\begin{figure*}
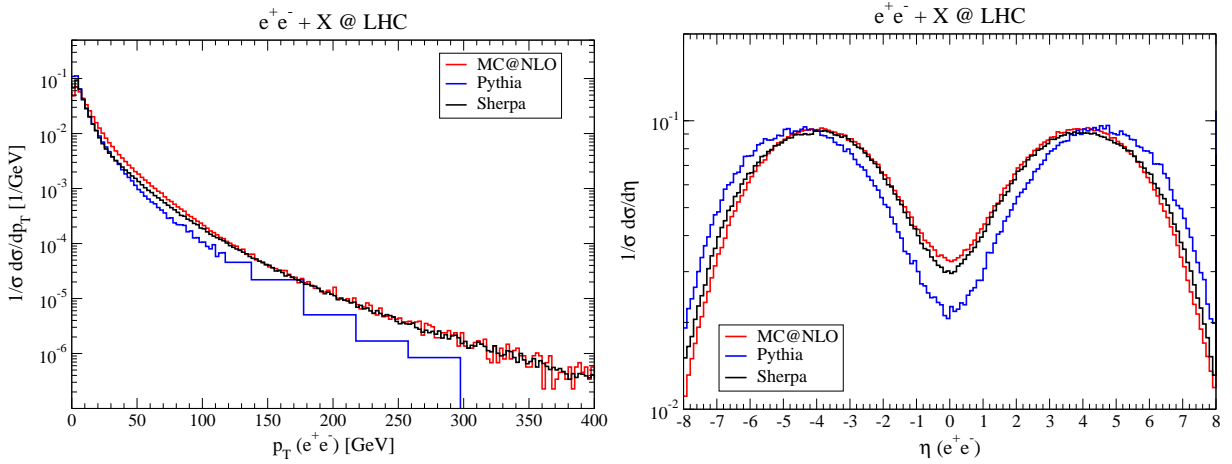

\begin{center}
\begin{tabular}{cc}
\includegraphics[width=8cm]{figures/LHC_Z_pt_MC.eps}&
\includegraphics[width=8cm]{figures/LHC_Z_eta_MC.eps}
\end{tabular}
\end{center}
\caption{\label{Z_pt_eta_MC}  The $p_\perp$ (left) and $\eta$ (right) distribution of the 
lepton pair in inclusive production of a Z/$\gamma^*$ boson decaying into $e^+e^-$ 
at the LHC. The results of the generators {\tt MC@NLO} (red), {\tt PYTHIA} (blue), 
and \sherpa\ (black) are compared.}
\end{figure*}

\begin{figure*}
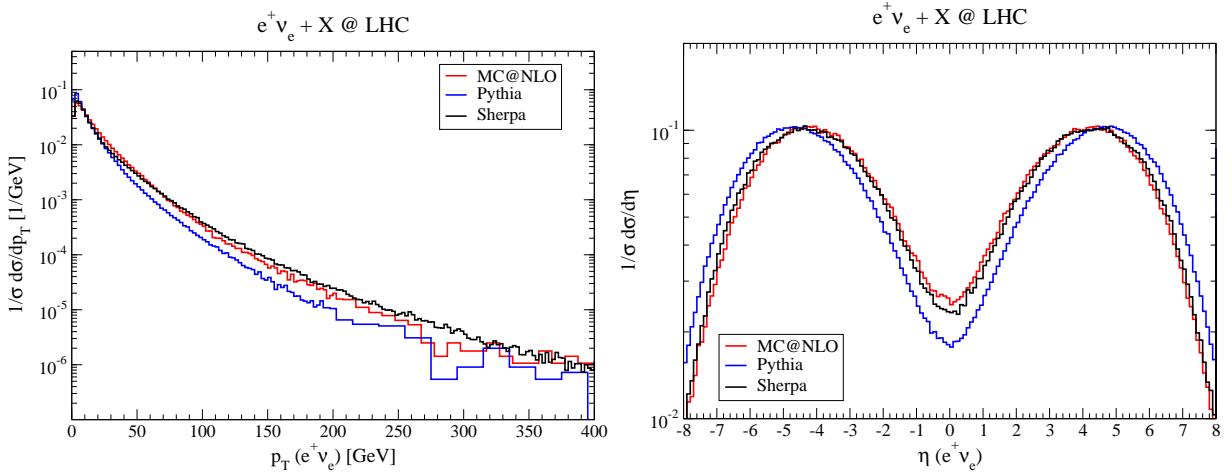

\begin{center}
\begin{tabular}{cc}
\includegraphics[width=8cm]{figures/LHC_Wp_pt_MC.eps}&
\includegraphics[width=8cm]{figures/LHC_Wp_eta_MC.eps}
\end{tabular}
\end{center}
\caption{\label{W_pt_eta_MC}  The $W^+$ $p_\perp$ (left) and $\eta$ (right) distribution 
in inclusive production at the LHC. The results of the generators {\tt MC@NLO} (red), {\tt PYTHIA} (blue), 
and \sherpa\ (black) are compared.}
\end{figure*}

\noindent
First of all, the results of the three programs for some rather inclusive quantities are 
compared. The transverse momentum and pseudo-rapidity distributions of the produced bosons 
are presented in Fig.\ \ref{Z_pt_eta_MC} and Fig.\ \ref{W_pt_eta_MC}. The \sherpa\ predictions 
depicted have been obtained with $n_{\rm max}=1$ and $Q_{\rm cut}=20$ GeV, in order to
match the approaches of the other codes. In order to compare the different samples, they all 
have been subject to a cut on the boson invariant mass of the form
\begin{align}\label{Vmass}
m_V-30\cdot\Gamma_V\le m_V^* \le m_V+30\cdot\Gamma_V\,,
\end{align}
where no additional phase space cuts have been applied. All distributions have been normalised
to their respective cross section.

\noindent
The results for both processes look very similar. The boson transverse momentum distributions of 
{\tt MC@NLO} and \sherpa\ agree fairly well. In the case of $Z/\gamma^*$ production they match 
nearly perfectly for values of $p_\perp> 100$ GeV. In the intermediate range of 
$10\,{\rm GeV} < p_\perp < 100\, {\rm GeV}$ \sherpa\ apparently is below {\tt MC@NLO}. This 
discrepancy may have its origin in the different shower approaches used within the two programs. 
This statement is also hinted at by the fact that the {\tt PYTHIA} result follows the \sherpa\ 
distribution for $p_\perp < 35$ GeV. For larger values of $p_\perp$, however, the {\tt PYTHIA} 
distribution is far below {\tt MC@NLO} and \sherpa\ predicting much less bosons with large 
transverse momentum. For the case of $W^+$ production the {\tt MC@NLO} and \sherpa\ prediction 
cross at $p_\perp \approx 60$ GeV. \sherpa\ produces slightly less events with smaller boson 
$p_\perp$ and tends to pronounce the high $p_\perp$ region a bit. Again {\tt PYTHIA} produces 
fewer bosons with intermediate and large boson transverse momenta. Looking at the pseudo-rapidity 
distributions, it can be recognized, that {\tt MC@NLO} and \sherpa\ both tend to produce the 
bosons much more central than {\tt PYTHIA}. Especially the region of $|\eta| < 4$ is filled 
significantly with respect to {\tt PYTHIA}, which, in contrast, features a much broader shape. 
This effect is of course directly correlated to the larger amount of hard QCD radiation the 
other two programs produce, since this enhanced QCD radiation allows for larger boson recoils. 
Moreover, from Fig.\ \ref{nmax_pt} it can be anticipated how the \sherpa\ results change under 
the inclusion of matrix elements with extra QCD legs: the boson transverse momentum distribution 
develops a more pronounced large $p_\perp$ tail and the very cental region of $\eta$ is filled 
even more, thus reducing the amount of events with large values of $|\eta|$. So while the 
$p_\perp$ spectra would be slightly harder than those of {\tt MC@NLO} the $\eta$ distributions 
would fit even better than they do for the case of including $V+0$ and $V+1$ parton matrix elements 
only.
 
\begin{figure*}
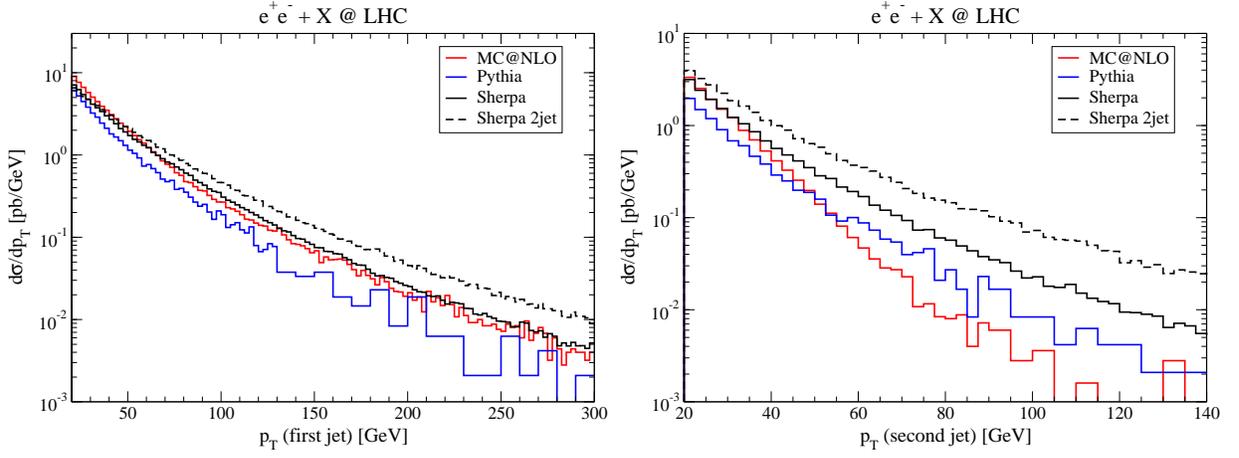

\begin{center}
\begin{tabular}{cc}
\includegraphics[width=8cm]{figures/LHC_Z_jet1_pt_MC_full.eps}&
\includegraphics[width=8cm]{figures/LHC_Z_jet2_pt_MC_full.eps}
\end{tabular}
\end{center}
\caption{\label{Z_jet12_pt_MC_full} The $p_\perp$ distribution of the first (left) and second (right) 
  hardest jet in inclusive $Z/\gamma^*$ production at the LHC as obtained by
  {\tt MC@NLO} (red), {\tt PYTHIA} (blue) and \sherpa\ (black). The dashed curves correspond 
  to the predictions of \sherpa\ when matrix elements for up to two additional partons are used.}
\end{figure*}

\begin{figure*}
\begin{center}
\includegraphics[width=8cm]{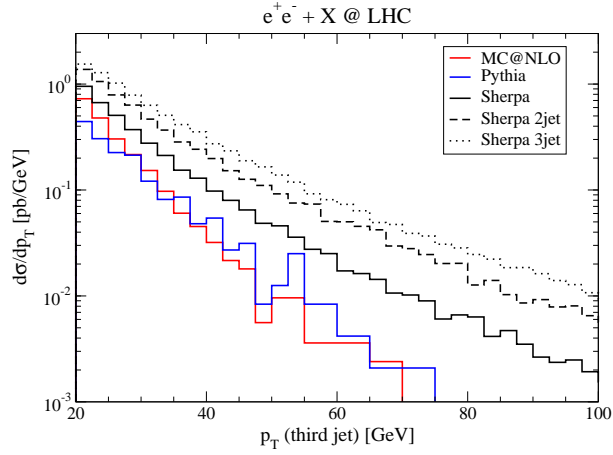}
\end{center}
\caption{\label{Z_jet3_pt_MC_full} The $p_\perp$ distribution of the third hardest jet in inclusive 
  $Z/\gamma^*$ production at the LHC as obtained by {\tt MC@NLO} (red), 
  {\tt PYTHIA} (blue) and \sherpa\ (black). The dashed (dotted) curve corresponds to the prediction 
  of \sherpa\ when matrix elements for up to two (three) additional partons are used. }
\end{figure*}

\noindent
For the comparison of jet observables, only the case of $Z/\gamma^*$ production is studied.
The qualitative statements implied by it, however, will hold true as well in the case of $W$ 
production. To judge the abilities of the three programs to produce extra hard QCD radiation 
associated to the electro-weak gauge bosons, the transverse momentum distributions of the three 
hardest jets are depicted in Fig.\ \ref{Z_jet12_pt_MC_full} and Fig.\ \ref{Z_jet3_pt_MC_full}. 
For this comparison in addition to the cut on the boson transverse mass according to 
Eq.\ (\ref{Vmass}) the jet criteria and phase space cuts of appendix \ref{appendix} have been 
applied. For \sherpa\ the jet resolution parameter has been set to $Q_{\rm cut}=20$ GeV. 
The standard sample for this comparison uses again only matrix elements with up to one 
additional parton. To test the predictions of \sherpa\ samples with $n_{\rm max}=2(3)$ 
have been considered as well, the corresponding results are shown as dashed (dotted) 
lines in the plots. Since it is actually the production rate that is important here, 
this time the curves have not been normalized. Instead the corresponding differential 
cross sections are presented.  

\noindent  
For the hardest jet the predictions of {\tt MC@NLO} and \sherpa\ agree rather well.
The total rate of \sherpa\ is $12\%$ smaller than that predicted by {\tt MC@NLO}, the 
distribution, however, has a slightly harder tail. This difference in rate can be 
traced back to the different inclusive production cross sections. 
However, for $n_{\rm max}=2$, the two total cross sections of $Z+1$jet nearly 
coincide (cf.\ the dashed black curve in Fig.\ \ref{Z_jet12_pt_MC_full}). In terms of 
shape, \sherpa\ apparently favours jets with larger transverse momentum. As has been seen 
in the closely related boson $p_\perp$ distribution in Fig.\ \ref{Z_pt_eta_MC}, 
{\tt PYTHIA} predicts a much smaller rate (60\% w.r.t.\ the rate predicted by 
{\tt MC@NLO}) for the production of extra hard QCD radiation with a softer distribution.

\noindent
For the second jet the situation changes significantly. Here, even in the case of including only 
matrix elements with up to one extra parton the two-jet rate predicted by \sherpa\ is $17\%$ larger 
than that of {\tt MC@NLO}. Including matrix elements with two extra partons the difference becomes 
nearly $90\%$. As for the case of the first jet, {\tt PYTHIA} predicts the radiation of a second jet 
with a much smaller rate. Similar statements hold true when looking at the third jet but this time the 
differences are even larger. Note, that a reliable prediction of the three jet rate requires the 
inclusion of matrix elements with at least three extra partons. While the sample with matrix elements 
up to one extra parton predicts a three jet rate of $9.6$ pb, the samples with two (three) extra partons 
predict $16.3\,(21.1)$ pb. However, this is not surprising keeping in mind that the LHC 
provides enough phase space to produce massive bosons in association with a multitude of high 
energetic jets that are best described by the corresponding matrix elements and that can not be
appropriately described by parton shower emissions.

\noindent
To summarize: the predictions of {\tt MC@NLO} and \sherpa\ agree fairly well for the shape of the boson 
transverse momentum and pseudo-rapidity distribution. Here, {\tt MC@NLO} is of course superior in 
predicting the rate of inclusive $Z/\gamma^*$ and $W$ production since it considers the corresponding 
production process at NLO in the coupling constant. This situation changes when studying the jets that 
potentially accompany the boson. As soon as more than one extra jet is considered \sherpa\ predicts 
significantly larger jet production rates and jet transverse momentum distributions that feature an 
enhanced population of the high-$p_\perp$ region. Concerning {\tt PYTHIA} it has to be stated that 
the shape of the boson transverse momentum and the boson pseudo-rapidity distribution differ 
significantly from the two other programs. This is directly related to the smaller amount of hard 
radiation produced by {\tt PYTHIA}, clearly observed in the jet $p_\perp$ spectra. 

\section{Summary}

\noindent
In this publication, the previous validation of the merging procedure of matrix elements and the parton 
shower, as implemented in \sherpa, has been continued. Again, processes of the type $pp\to V+X$, where 
$V = W^\pm,\,Z/\gamma^*$ have been chosen; this time, however, the analysis focused on the case of the 
CERN LHC rather than on the Fermilab Tevatron. Again, the merging procedure turned out to yield 
sufficiently stable results over a wide range of internal parameters, rendering it a predictive way of 
incorporating the full information available in tree-level matrix elements into multi-purpose event 
generators, as anticipated. In addition, when comparing the results obtained through \sherpa\ with 
those of a full next-to-leading order calculation, it again turned out that the results of 
\sherpa\ reproduce the essential features in the NLO shapes. However, it should be stressed 
that in \sherpa\ the total rates are still at leading order accuracy only. Nevertheless, the fact 
that \sherpa\ seems to reproduce the NLO shapes of the observables, the NLO rates can be recovered 
by simply multiplying with a constant $K$-factor. When comparing the results of \sherpa\ with 
those of other event generators, some differences appear. Especially for observables sensitive 
to the correct treatment of multi-particle final states these differences have become significant, 
ranging up to orders of magnitude.  

\noindent
In this study, \sherpa\ again proved its versatility in simulating high-multiplicity final states 
at collider experiments. Due to the merging procedure implemented in it, it provides a unique tool
for the simulation of final states, where the proper treatment of the event topology is of great
importance.

\section*{Acknowledgments}
\noindent
The authors would like to thank T.\ Gleisberg, S.\ H{\"o}che, and J.\ Winter for pleasant
collaboration on \sherpa\ and M.\ L.\ Mangano and J.\ Campbell for helpful discussions. Financial
support by BMBF, DESY-PT, GSI, and DFG is gratefully acknowledged. 

\appendix
\section{Input parameters and phase-space cuts}\label{appendix}
\noindent
For all analyses, the PDF set cteq6l \cite{Pumplin:2002vw} has been used, and $\alpha_S$ has 
been chosen according to the corresponding value of this PDF, namely $\alpha_S = 0.118$. 
For the running of the strong coupling constant, the corresponding two-loop equation has 
been employed. Jets or initial partons are restricted to the light flavor sector, namely 
$g,\,u,\,d,\,s,$ and $c$. All flavors are taken to be massless.
\subsection{SM input parameters}
\noindent
The SM parameters are given in the $G_{\mu}$ scheme:
\begin{eqnarray}
&&m_W = 80.419\;{\rm GeV}\,,\quad \Gamma_W = 2.06\;{\rm GeV,}\nonumber\\
&&m_Z = 91.188\;{\rm GeV}\,,\quad \Gamma_Z = 2.49\; {\rm GeV,}\nonumber\\
&&G_{\mu} = 1.16639 \times 10^{-5}\; {\rm GeV}^{-2},\nonumber\\
&&\sin^2\theta_W = 1 - m^2_W/m^2_Z,\nonumber\\
&&\alpha_s = 0.118.
\end{eqnarray}
\noindent
The electromagnetic coupling is derived from the Fermi constant $G_{\mu}$
according to
\begin{eqnarray}
&&\alpha_{\rm em} = \frac{\sqrt{2}\,G_{\mu}\,M^2_W\,\sin^2\theta_W}{\pi}\,.
\end{eqnarray}
\noindent
The constant widths of the electroweak gauge bosons are introduced through the 
fixed-width scheme. CKM mixing of the quark generations is neglected.

\subsection{Cuts and jet criteria}

\noindent
For all jet analysis the Run II $k_\perp$ clustering algorithm defined in \cite{Blazey:2000qt} 
has been used. The additional parameter of this jet algorithm is a pseudo-cone size $R$, whose 
value has been chosen to $R=0.4$. In addition jets have to fulfil the following cuts on 
transverse momentum and pseudo-rapidity, 
\begin{eqnarray}
p_T^{\rm jet} > 20\; {\rm GeV}, \quad |\eta^{\rm jet}| < 4.5.
\end{eqnarray}

\noindent
For charged leptons the cuts applied are:
\begin{eqnarray}
p^{\rm lepton}_T > 15\;{\rm GeV},\quad |\eta^{\rm lepton}| < 2.4,\quad m_{ll} > 15\; {\rm GeV}.
\end{eqnarray}
No cut on missing transverse momentum has been applied.

\noindent
The final selection criteria correspond to the separation of the leptons 
amongst each other and with respect to the jets,
\begin{eqnarray}
\Delta R_{lj} > 0.4\; , \quad \Delta R_{ll} > 0.2.
\end{eqnarray}

\section{Setups for MC@NLO and PYTHIA}\label{appendix_mc}
\begin{itemize}
\item {\underline{The {\tt MC@NLO} setup:}}
  The program version used is {\tt MC@NLO 2.31}. The process number corresponding 
  to $pp \to Z/\gamma^* \to e^+e^- + X$ production is {\tt IPROC=-11351}, for 
  $pp \to W^+ \to e^+\nu_e + X$ this equals {\tt IPROC=-11461}.
  In both cases consequently the underlying event has been switched off. 
  The lepton pair in $Z/\gamma^*$ production has been generated in a mass window of 
  \begin{eqnarray}\label{Zmass}
    m_Z-30\Gamma_Z \leq m_{ee} \leq m_Z+30\Gamma_Z
  \end{eqnarray}
  in the case of $W^+$ production the lepton-neutrino pair fulfils 
  \begin{eqnarray}\label{Wmass}
    m_W-30\Gamma_W \leq m_{e\nu_e} \leq m_W+30\Gamma_W
  \end{eqnarray}
  The PDF set used is cteq6l. All other physics parameters that specify a run for {\tt MC@NLO} 
  have been left unchanged with respect to their default values.  
\item {\underline{The {\tt PYTHIA} setup:}} 
  The {\tt PYTHIA} version used is {\tt 6.214}. The process $pp \to Z/\gamma^* + X$ 
  is selected via the parameter {\tt MSUB(1)=1}. The decay mode 
  $Z/\gamma^* \to e^+e^-$ is picked by the settings {\tt MDME(182,1)=1} and {\tt MDME(170,1)=1}, 
  all other decay channels have been disabled. The process $pp \to W^+ + X$ is 
  turned on via ${\tt MSUB(2)=1}$. The decay mode $W^+ \to e^+\nu_e$ is chosen by {\tt MDME(206,1)=1}. 
  It has proven to be convenient to increase the standard value of the shower starting scale in 
  {\tt PYTHIA} to $\sqrt{s}=14$ TeV in order to produce a reasonable amount of high energetic 
  QCD radiation. The parameter responsible for controlling the shower start scales is 
  {\tt MSTP(68)}. For the above choice it has set to set to {\tt MSTP(68)=2}. 
  All other parameters specific for {\tt PYTHIA} have been left unaltered, except that 
  {\tt PYTHIA}'s underlying event has been switched off.  
\end{itemize}


\end{document}